\documentclass{sig-alternate}
\usepackage{amssymb}
\usepackage{mathrsfs}
\usepackage{amsfonts}

\usepackage{graphicx}
\usepackage{subfigure}
\usepackage{algorithm}
\usepackage{algorithmic}
\usepackage{amsmath}
\usepackage{multirow}

\newcommand{\fullversion}[1]{}

\begin{document}

%

\title{News-Based Group Modeling and Forecasting}
%
%
%
%
%

\numberofauthors{2} 
%

\author{
%
%
\alignauthor
Wenbin Zhang\\
       \affaddr{Department of Computer Science}\\
       \affaddr{Stony Brook University}\\
       \affaddr{Stony Brook, NY 11794-4400 USA}\\
       \email{wbzhang@cs.sunysb.edu}
\alignauthor
Steven Skiena\\
       \affaddr{Department of Computer Science}\\
       \affaddr{Stony Brook University}\\
       \affaddr{Stony Brook, NY 11794-4400 USA}\\
       \email{skiena@cs.sunysb.edu}
}

\maketitle

\begin{abstract}
In this paper, we study news group modeling and forecasting methods
using quantitative data generated by our large-scale natural
language processing (NLP) text analysis system. A news group is a
set of news entities, like top U.S. cities, governors, senators,
golfers, or movie actors. Our fame distribution analysis of news
groups shows that log-normal and power-law distributions generally
could describe news groups in many aspects. We use several real news
groups including cities, politicians, and CS professors, to evaluate
our news group models in terms of time series data distribution
analysis, group-fame probability analysis, and fame-changing
analysis over long time. We also build a practical news generation
model using a HMM (Hidden Markov Model) based approach. Most
importantly, our analysis shows the future entity fame distribution
has a power-law tail. That is, only a small number of news entities
in a group could become famous in the future. Based on these
analysis we are able to answer some interesting forecasting problems
- for example, what is the future average fame (or maximum fame) of
a specific news group? And what is the probability that some news
entity become very famous within a certain future time range?  We
also give concrete examples to illustrate our forecasting
approaches.

\end{abstract}




\section{Introduction}

You will never see a newspaper headline announcing that the sun came
up yesterday. This is because news, by definition, must be
unpredictable: reporting on unexpected events around the world.
Attempting to predict the contents of tomorrow's news seems a
misguided and futile task. And yet news prediction is regularly
attempted in several domains, including financial modeling, weather
forecasting, and political polling.

In this paper, we applying modeling and forecasting techniques to
the future reference frequency of people, places, and things in the
news. We seek to estimate the probability that a given entity (or
set of entities) $E$ will be mentioned in the news at least $f$
times over the next time period $t$.  Our techniques are analogous
to volatility-based financial models which attempt to predict the
probably future trading range of a given stock, as opposed to the
unknowable question of whether it will go up or down tomorrow. Our
news forecasting methods can be used to answer questions like:

\begin{list}{\labelitemi}{\leftmargin=1.5em} \itemsep -1pt
\item
What are the chances a particular political party will suffer a
significant scandal over the next year?
\item
Will any other celebrity death over the next decade attract the same
media coverage as Michael Jackson's?
\item
How famous will the most successful graduate of your college class
become?
\end{list}

The Lydia system (\cite{Lloyd05}, http://www.textmap.com), a project
developed in the Algorithms Lab at Stony Brook University, is
capable of capturing quantitative news time series, and analyzing
spatial, temporal, and linguistic statistics of named entity
occurrences over a large corpus of news text. This makes Lydia data
a perfect source to analyze daily news with respect to our
time-series-world. \fullversion{News data analysis is able to
provide interesting views for the study of financial analysis,
political science, and even other fields of social science. If we
can understand what happened in the news, we should be able to
understand the past and perhaps predict the future.}

Lydia system identifies news entities or synonymous sets (synsets),
and provides their daily statistics in terms of their frequency,
sentence counts, article counts, and sentiment counts. News entities
mean entity names like ``Tiger Woods", ``Summer Olympics", or
``Lehman Brothers", while entities ``Lehman Brothers Holdings" and
``Lehman Brothers Inc." refer to the same synset ``Lehman Brothers".
Lydia system analyzes and tracks all entities occurring in several
different depositories, among which the {\em Dailies} depository has
the biggest data volume and thus it will be used for our analysis.
Indeed, the Dailies depository constructs the entity/synset time
series from over one terabyte of U.S. and international
English-language newspapers, starting in November 2004. It contains
news from about 500 different sources each day.

Actually, most interesting forecasting problems are considered under
the context of news groups. For example, top U.S. cities, governors,
college athletes, golfers, or movie actors are all groups. Our
purpose in this paper is to investigate news group models, find out
news group data generation rules underneath, and then build methods
to forecast the future of news groups. For example, can we build
model to forecast the future average fame (or maximum fame) of a
specific group? And can we predict the probability that some entity
in a group become famous within a certain time frame? These topics
are intensively related to the area of data modeling and
forecasting. However, to the best of our knowledge, these particular
news-based forecasting problems have never been seriously studied
before. More specifically, Our contributions in this paper are:

\begin{list}{\labelitemi}{\leftmargin=1.5em} \itemsep -1pt

\item
{\em Statistical Modeling on the Emergence of Fame} -- Through
extensive computation on our terabyte-scale news corpus, we study
changes in the reference frequency among various classes of
entities. Future reference frequency can be modeled as a combination
of log-normal distributions (for frequent entities) and power law
distributions (for less frequently mentioned ones), captured using
an appropriate hidden Markov model (HMM).

\item
{\em Group Frequency Analysis} -- Predicting phenomena like the
frequency of political scandals requires forecasting the future of
large groups of individuals. We generalize our forecasting models to
answer questions on the total and maximum news volume among members
of a group.

\item
{\em Domain-specific News Forecasting} -- We apply our news
forecasting techniques to three interesting domains with different
sizes (Top 50 U.S. cities, representatives, and computer science
faculty) and backtest these models over historical news data to
confirm the general validity of our models.

\end{list}

The contents of this paper are organized as follows. We review
related work in Section \ref{previousWork}. Section
\ref{statisticalProperties} studies the statistical patterns for
news entities and groups. In Section \ref{hmmBased}, we propose a
HMM based news generation model and evaluate its accuracy. We build
models to solve entity fame and domain-specific fame forecasting
problems and validate them in Section \ref{entityForecasting} and
\ref{groupForecasting}. We give conclusions in Section
\ref{conclusions}.

\section{Previous Work} \label{previousWork}

Our work here is related to several existed research directions:
 news frequency modeling, news event/topic modeling, and modeling
method for other relevant data streams.

Leskovec \cite{Leskovec09} focuses on the study of the dynamics of
news cycle, i.e., to model the process of news start, reaching
peaks, and decay. The authors use a so-called ``meme-tracking"
method to track short, distinctive phrases through on-line text, and
show that this method is capable of tracking information spread over
Internet and providing a coherent representation of news cycle. They
also developed a mathematical model to describe the trend of news
cycle, in which both the imitation effects and recency effects of
news sources are considered. Although the mathematical news model is
proposed, the authors have neither tried to fit the model with real
data to validate this model, nor shown the goodness of this model to
predict future news.

Other work focuses on the modeling of some other data streams, like
blog behavior, disk I/O traffic, or network traffic data. Gotz et
al. \cite{Gotz2009blog} studied the temporal behavior in
blogosphere, and proposed a $\mathcal {ZC}$ model to simulate blog
behavior, which uses a `zero-crossing' approach based on random
walk. Wang et al. \cite{Wang02} proposed a $b$-model for disk I/O
traffic data simulation, which is a good fit for self-similar data
traffic. The authors also provided a fast algorithm to implement the
$b$-model. Leland et al. \cite{Leland94} also analyzed the
self-similarity of Ethernet traffic data based on statistical
analysis and discussed the significance of self-similarity. Johnson
et al. \cite{Johnson06} reported the power-law behavior behind
terrorist attacks and wars.

In the topic modeling area, all the research works are somehow
derived from Latent Semantic Analysis (LSA) technique, which was
first introduced in U.S. patent \cite{Deerwester89}. LSA technique
analyzes relationships between documents and the terms they contain,
and then generate a set of concepts which are related to the
documents and terms. Research which can characterized under this
model includes \cite{Hofmann99}, \cite{Blei03}, \cite{Viet09}.
However, we should note that all LSA related research focuses on
recognizing news topics, and these techniques are not quite relevant
to forecast future topic trend.

There are several commercial products which track events or topics
from News or Blogs, like Google Trend \cite{GoogleTrends}, Google
Insight Search \cite{GoogleInsight}, Blog Pulse \cite{BlogPulse},
and Blog Scope \cite{BlogScope}. Basically, they only monitor events
from News or Blogs, but they are not relevant to new group modeling
or predictions.

\section{Statistical Properties of News Entities and Groups} \label{statisticalProperties}

Objects we need to study include news entities and groups. A group
is a set of news entities with certain common attributes. Table
\ref{groups} shows the news groups used in following sections. These
groups fall into three categories according to their group sizes,
i.e., small, medium, and large groups respectively.

\begin{table}[pht]
\scriptsize \centerline{
\begin{tabular}
{l||r|l} \hline Group & Size & Description
\\ \hline \hline
Africa & 51 & Countries in Africa \\
Top 50 US Cities & 50 & The top 50 U.S. cities (by population) \\
Governors &  49 & Current United States governors \\ \hline
Senators & 105 &Current United States senators \\
Representatives & 439& Current United States representatives \\
NCAA/Big Ten & 275 & Players in Big Ten Conference in NCAA
\\ \hline CS Professors & 1911 & CS Professors in Top
40 CS
departments \\
Golfers & 1749 & The completed list of Golfers\\
Football Players & 2255 & Current National Football League Players\\
Hockey Players & 5986 &  All National Hockey League (NHL) players\\
Movie Actors & 47146 & Actors who performed movies in 2000-2010\\
 \hline
\end{tabular}}
\caption{Some News groups and their descriptions. \label{groups}}
\end{table}

\subsection{Fame and Fame Window}

\begin{table*}[htp]
\scriptsize \centerline{
\begin{tabular}{
l||r|r|r|r|r|r||r|r|r|r|r|r} \hline \multirow{3}{*}{Entities}&
\multicolumn{6}{c||}{Raw Frequencies (Fame)} &
\multicolumn{6}{c}{Logged Frequencies (Fame)} \\
\cline{2-13} & \multicolumn{3}{c|}{Average Fame}&
\multicolumn{3}{c||}{Peak Fame } &
\multicolumn{3}{c|}{Average Fame}& \multicolumn{3}{c}{Peak Fame} \\
\cline{2-13} & 2007 &2008&2009&2007&2008&2009&2007&2008&2009&2007&2008&2009 \\
\hline \hline
United States & 6036&9262&7720&14041&15965&17889&8.705&9.134&8.952&9.550&9.678&9.792 \\ Barack Obama&2537&38598&33709&9869&129108&98844&7.839&10.561&10.426&9.197&11.768&11.501 \\
Chicago, IL & 2594 & 4567 & 3408&6228& 7429& 6380& 7.861&
8.426&8.134& 8.737& 8.913& 8.761 \\
Tiger Woods &647&1070&2064&4361&12984&20202&6.474&6.977&7.633&8.381&9.472 & 9.914\\
Michael Jackson & 424& 689 &8224 &1514&5317&123739&6.052&6.537&9.015&7.323&8.579& 11.723 \\
Steve Jobs &77 & 83&110&455&646&1055&4.365&4.435&4.706&6.123&6.472&6.962\\
George Clooney &0.963&1.402&1.100&14.200&18.999&35.800&0.675&0.876&0.742&2.721&2.996&3.605\\
Stephen Leeb &0.263&1.690&0.939&7.999&84.600&42.400&0.234&0.989&0.662&2.197&4.449&3.770\\
 \hline
\end{tabular}}
\caption{Fame examples for selected entities, including raw
frequencies and logarithmic frequencies. Average fame takes the
average daily frequencies for a certain year, while peak fame takes
the maximum fame for that particular year, with a certain fame
window size. Here the window size for peak fame calculation is
5-day. \label{entityFame}}
\end{table*}

\fullversion{
\begin{table*}[pht]
\scriptsize \centerline{
\begin{tabular}
{l||r|r|r|r|l} \hline Group & Size &Tot Freq & Max Freq & Ave Freq&
Description
\\ \hline \hline
Africa & 51 &&&& Countries in Africa \\
Top 50 U.S. Cities & 50 &&&& The top 50 U.S. cities ranked by population \\
Governors &  49 &&&& Current United States governors \\ \hline
Senators & 105 &&&& Current United States senators \\
Representatives & 439&&&& Current United States representatives \\
NCAA - America East & 275 &&&& Players in America East Conference in
NCAA (2005-2009) \\ \hline CS Professors & 1911 &&&& Computer
Science Professors in Top 40 CS
departments \\
Golfers & 1749 &&&& The completed list of Golfers\\
Football Players & 2255 &&&& List of current National Football League (NFL) Players\\
Hockey Players & 5986 &&&& The completed list of National Hockey League (NHL) players\\
Movie Actors & 47146 &&&& All actors who performed movies from 2000 to 2010\\
 \hline
\end{tabular}}
\caption{Some News groups and their descriptions. These news groups
are divided into three categories: small, medium, and large groups
respectively. \label{groups}}
\end{table*}
}

Entities' magnitudes (fame) differ significantly. A very common
entity like ``New York" may be mentioned everyday, while many other
entities may not. Fame is a term to measure an entity's magnitude in
a certain time period, and it could be measured by the average daily
references or logarithmic daily references. Based on an entity $E$'s
historical time series, the fame of it could be measured as below:
\begin{equation}
\displaystyle F_{E, N}=F(E, N)=log(1+\frac{\sum_{i=0}^{N-1}
{f_i}}{N}) \label{fame}
\end{equation}
In this formula, fame window size $N$ is the length of observation
window, usually measured in days, while $f_i$ is entity $E$'s
frequency counts of day $i$. Fame is actually also time series data,
and keeps changing over time. If the fame window size is 1, the fame
time series is just equivalent to the entity's daily frequency time
series.

Table \ref{entityFame} provides fame examples for selective
entities. It shows some popular entities like ``United States" or
``Barack Obama" as well as some unpopular entities like ``George
Clooney" or ``Stephen Leeb". In fact, ``United States" is one of the
top 10 entities in our news depository, while George Clooney (an
American actor) and Stephen Leeb (a computer scientist)  are very
trivial and they do not have much fame.

\subsection{Entity and Group Fame Distribution}
\label{distribution}

Now we focus on the daily fame distribution of news entities or news
groups. For any group $G$ (or entity $E$ with regular mentions), we
propose a log-normal distribution to model their daily fame
distribution $F(G, w_f)$ (or $F(E, w_f$)). Here the $w_f$ is fame
window size. For example, Figure \ref{hist-puerto} shows the daily
frequency distribution of Puerto Rico, in which the red curve shows
how good it fits to a standard normal distribution. This plot proves
log-normal distributions are good approximations for entity fame.
Similarly, we have exactly the same result for group fame. In fact,
group fame is the union of the fame of all entities in this group.

The log-normality of daily fame could be explained by {\em
multiplicative processes} \cite{Mitzenmacher_abrief}. Suppose we
start with news reference $F_0$. For each step $i$, the news
frequency may increase or decrease, then we have $F_i =
X_i{F_{i-1}}$, in which $X_i$ is a random variable. Therefore, we
can get
\begin{equation}
ln F_i = ln F_0 + \sum_{k=1}^{i} {ln X_k} \label{lognormal}
\end{equation}
$X_k$ are random variables. According to the Central Limit Theorem,
the $\sum_{k=1}^{i} {ln X_k}$ converges to a normal distribution.
Therefore, if $i$ is large, $F_i$ approximately follows a log-normal
distribution.

Especially, for a news group, the total fame, the maximum fame, and
the average fame all follow log-normal distributions according to
our analysis. \fullversion{Actually, a log-normal model is
equivalent to a geometric Brownian motion (GBM), which could be used
in our news generation model later.}

\begin{figure*}[ht]
\begin{minipage}[t]{0.29\linewidth}
\centering
\includegraphics[height = 1.5in, width=0.8\textwidth]{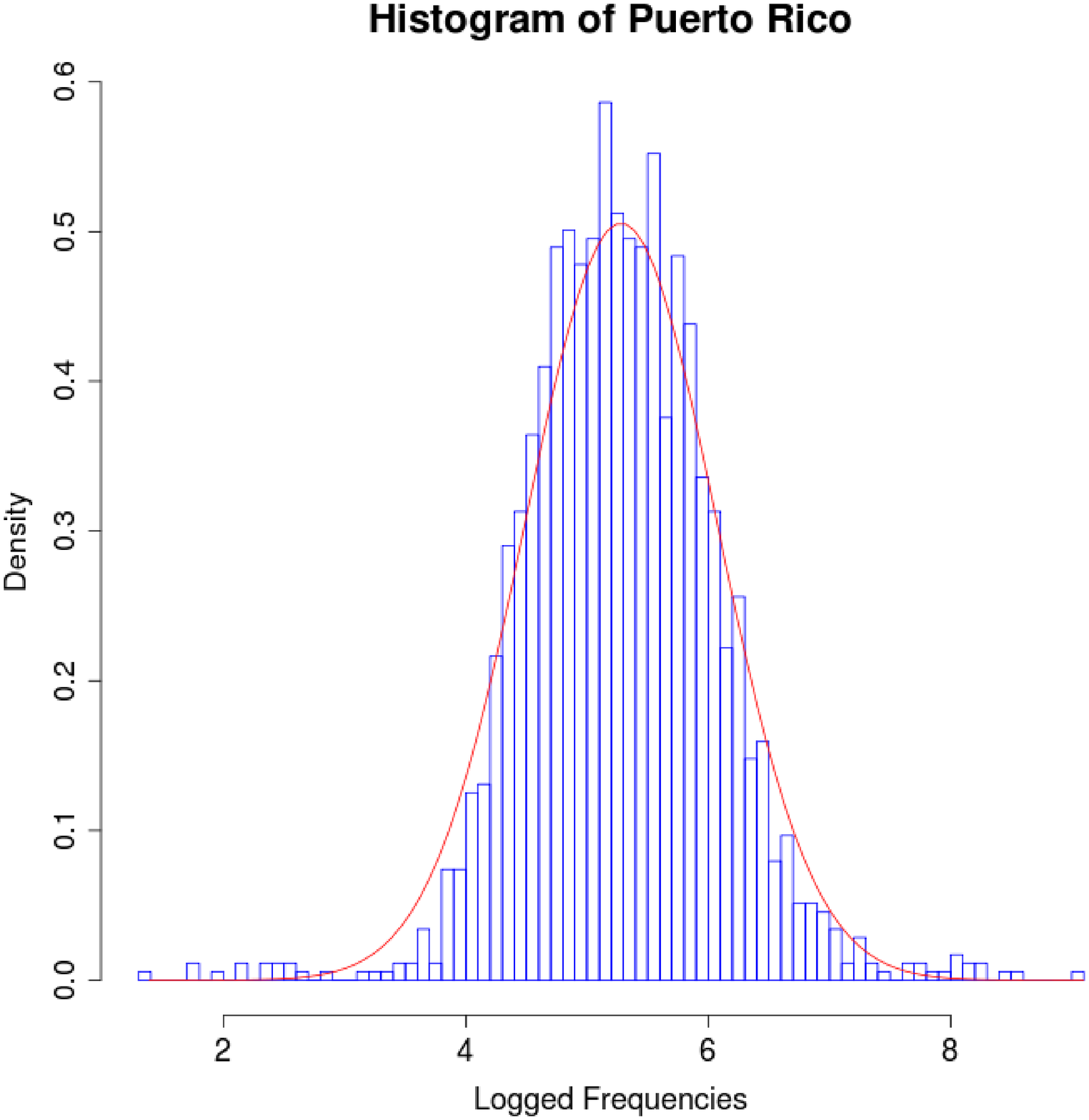}
\caption{Histogram plot for log-frequency of Puerto
Rico.\label{hist-puerto}}
\end{minipage}
\hfill
\begin{minipage}[t]{0.29\linewidth}
\centering
\includegraphics[height = 1.5in, width=0.8\textwidth]{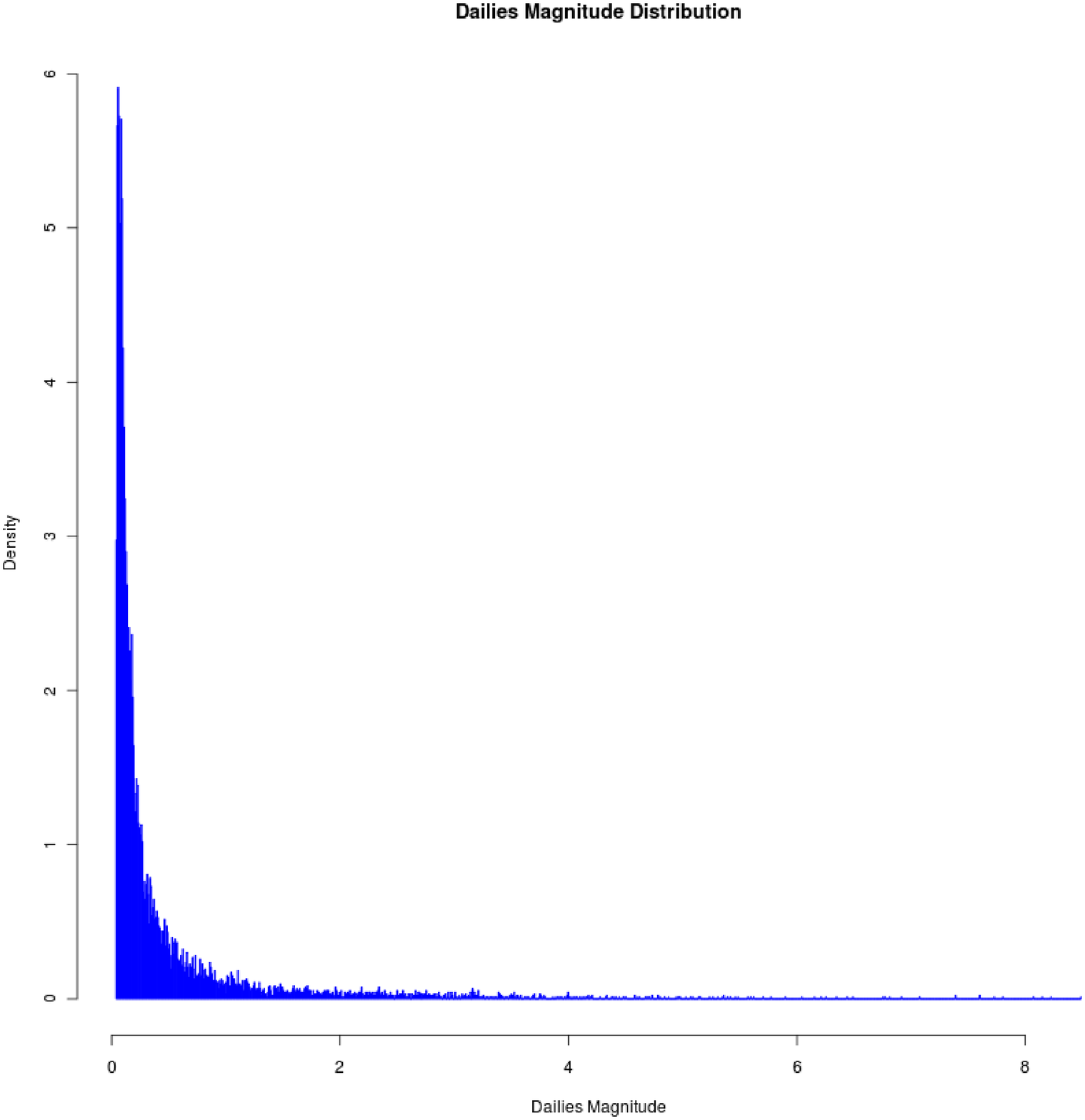}
\caption{Fame distribution of Dailies depository.
\label{dailies-dist}}
\end{minipage}
\hfill
\begin{minipage}[t]{0.39\linewidth}
\centering
\includegraphics[height = 1.6in,width=0.95\textwidth]{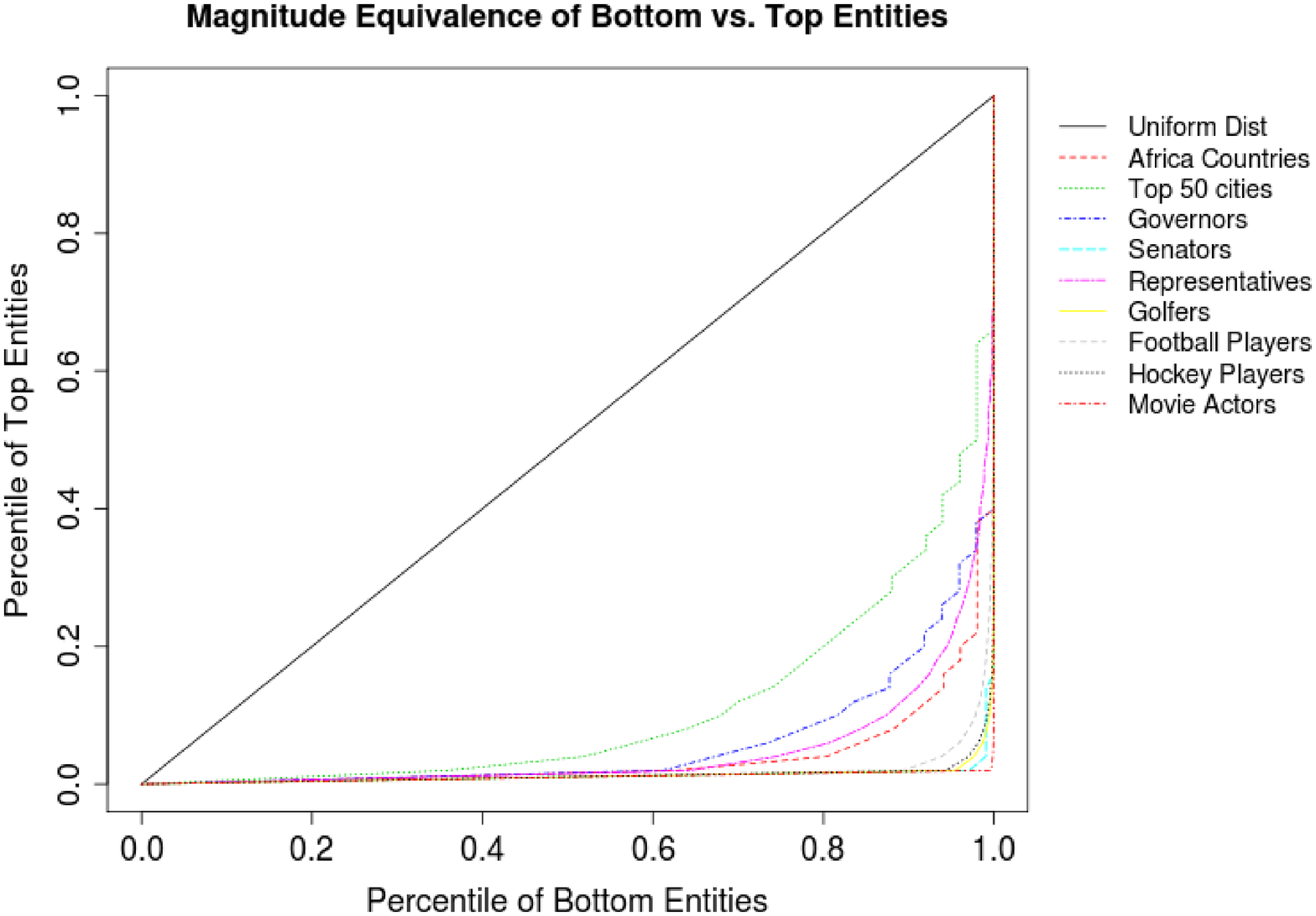}
\caption{Magnitude equivalence of bottom vs. top entities for
selective groups. \label{allEquiv}}
\end{minipage}
\end{figure*}

\fullversion{
\begin{figure*}[ht]
\begin{minipage}[t]{0.32\linewidth}
\centering
\includegraphics[height = 1.3in, width=0.7\textwidth]{./figures/add/hist-puerto.eps}
\caption{Histogram plot for logged frequency of Puerto Rico over 5
years.\label{hist-puerto}}
\end{minipage}
\hfill
\begin{minipage}[t]{0.32\linewidth}
\centering
\includegraphics[height = 1.3in, width=0.7\textwidth]{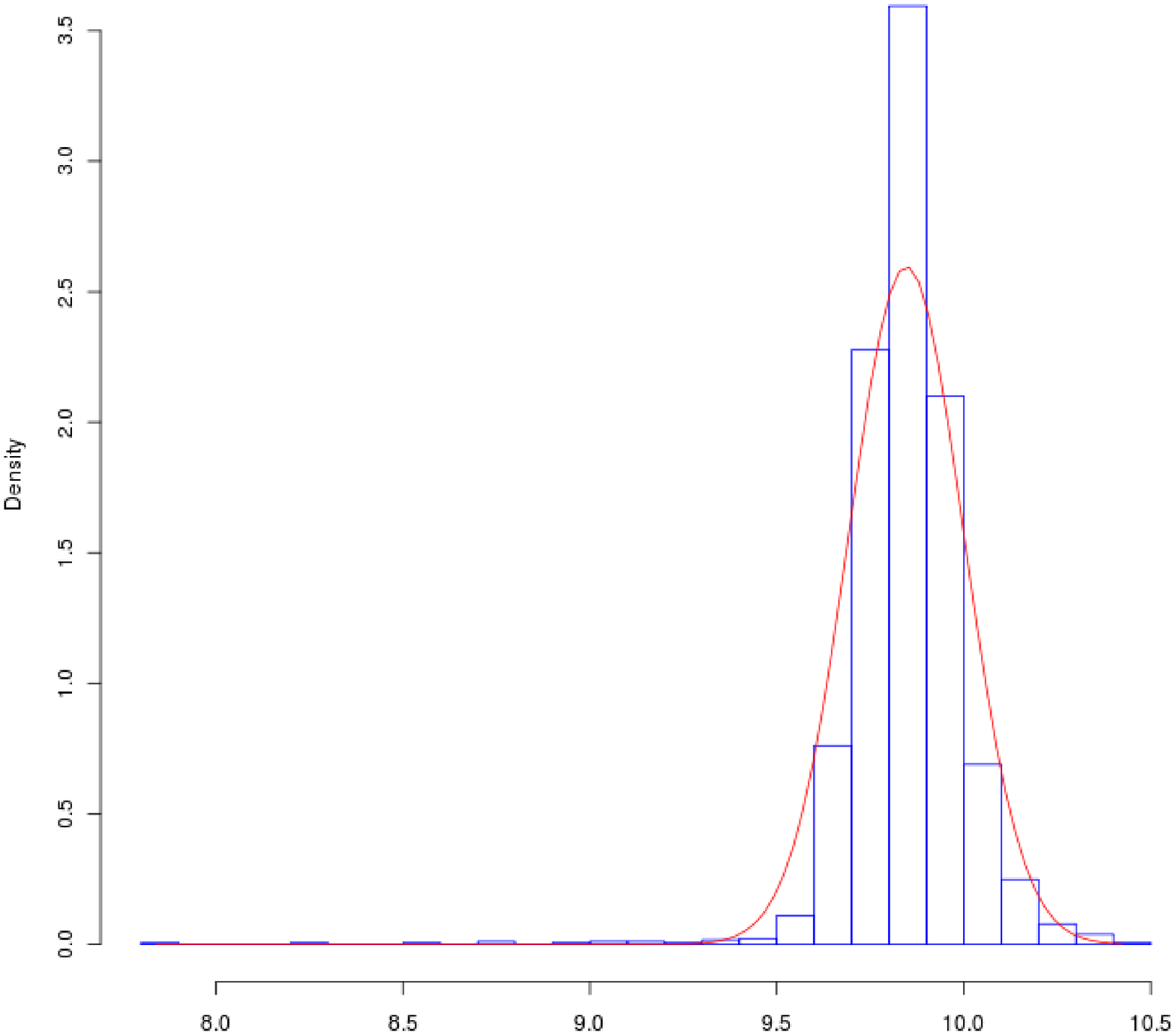}
\caption{Histogram plot for the total fame of {\em Top 50 U.S.
Cities}. The fame window size is 1 month. \label{cities-total-hist}}
\end{minipage}
\hfill
\begin{minipage}[t]{0.32\linewidth}
\centering
\includegraphics[height = 1.3in,width=0.7\textwidth]{./figures/dailies-dist.eps}
\caption{Fame distribution of Dailies
depository.\label{dailies-dist}}
\end{minipage}
\end{figure*}
}

\fullversion{
\begin{figure}[pht]
\begin{center}
\includegraphics[width=0.7\linewidth]{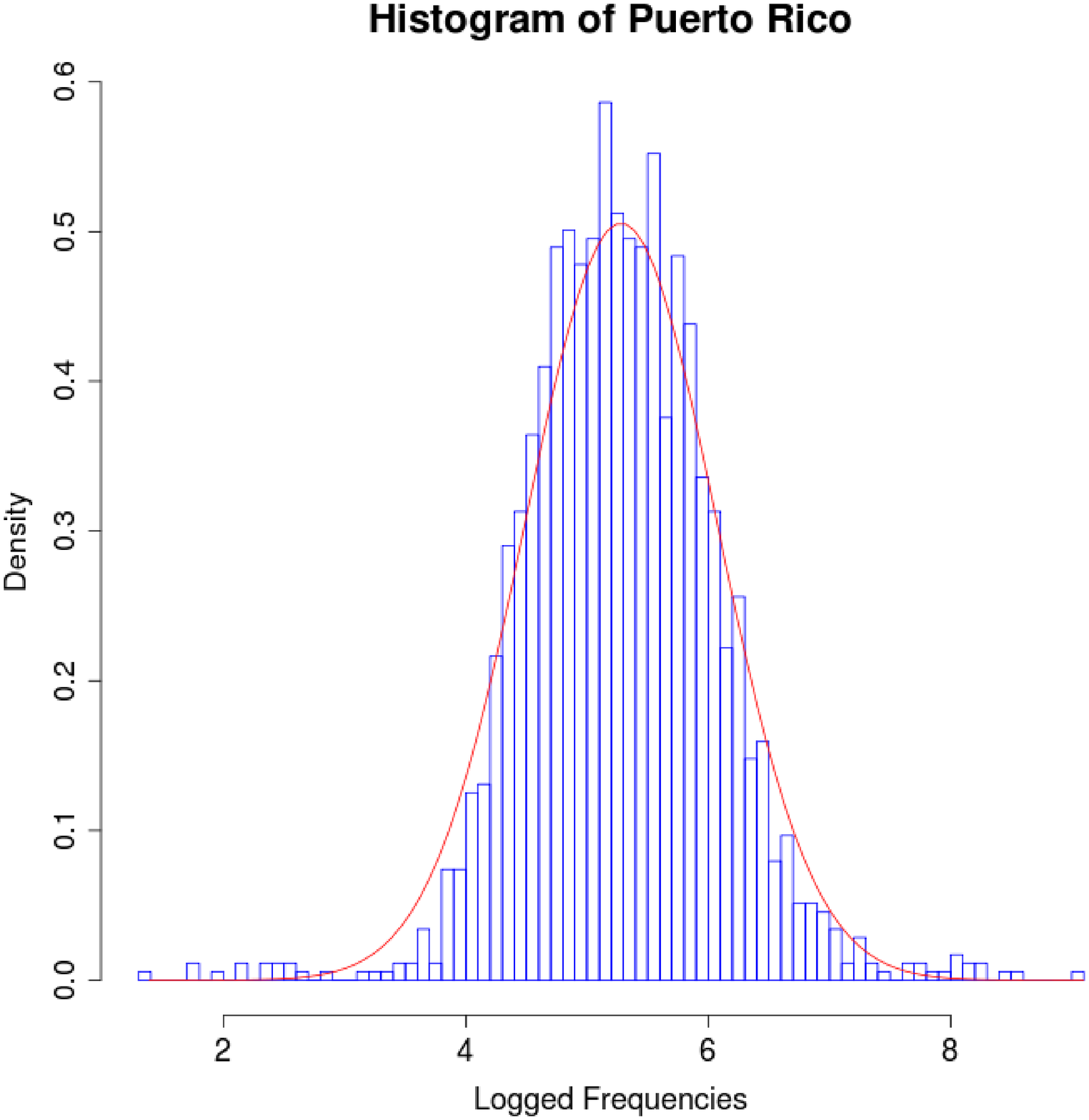}
\end{center}
\caption{Histogram plot for logged frequency of Puerto Rico over 5
years. The central of the plot is about the Puerto Rico's magnitude,
say, 5.6.} \label{cities_hist}
\end{figure}

\begin{figure}[pht]
\begin{center}
\includegraphics[width=0.7\linewidth]{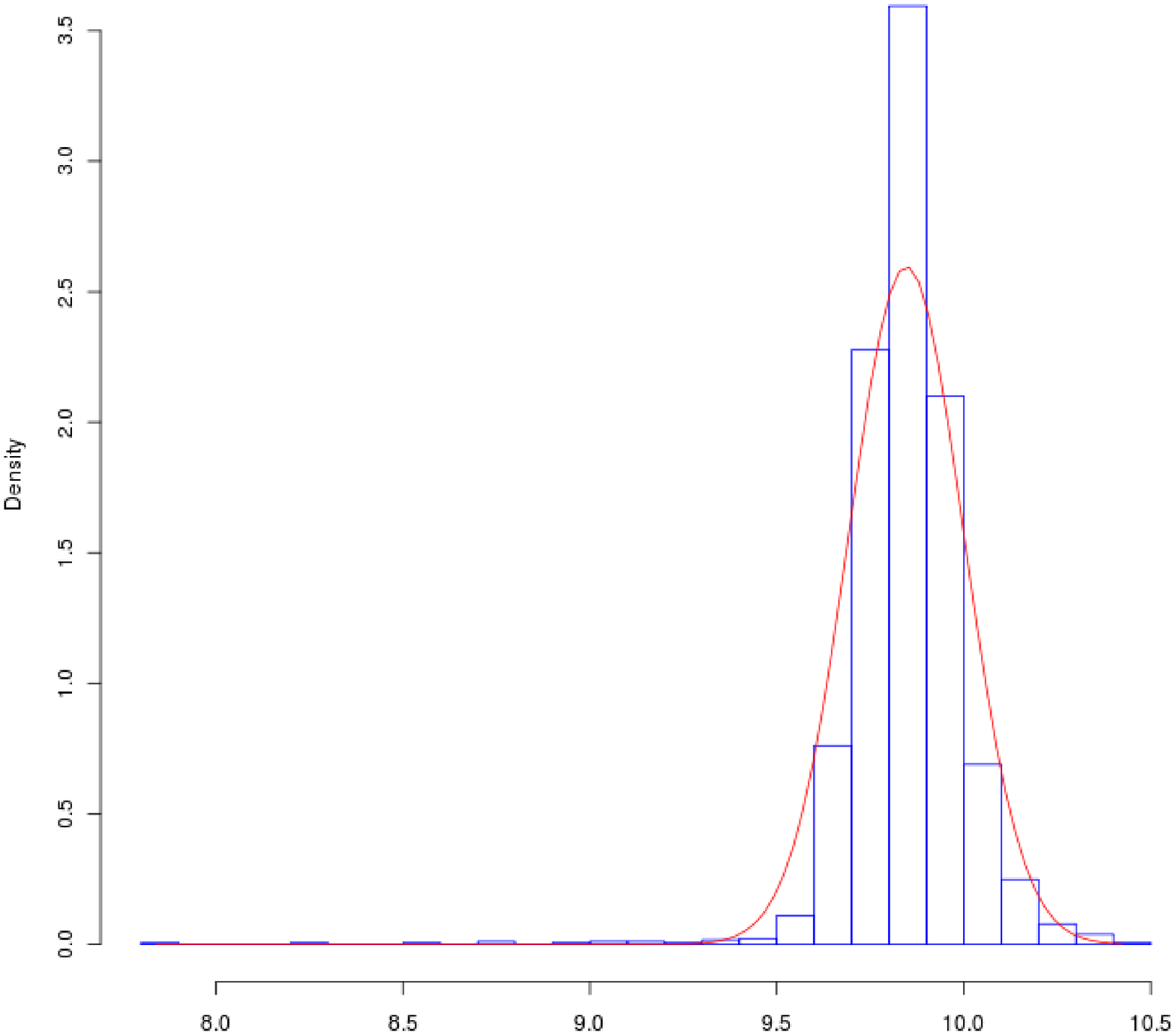}
\end{center}
\caption{Histogram plot for the total fame of group {\em Top 50 U.S.
Cities}. The fame window size is 1 month.} \label{cities_hist}
\end{figure}

\begin{figure}[pht]
\begin{center}
\includegraphics[width=0.7\linewidth]{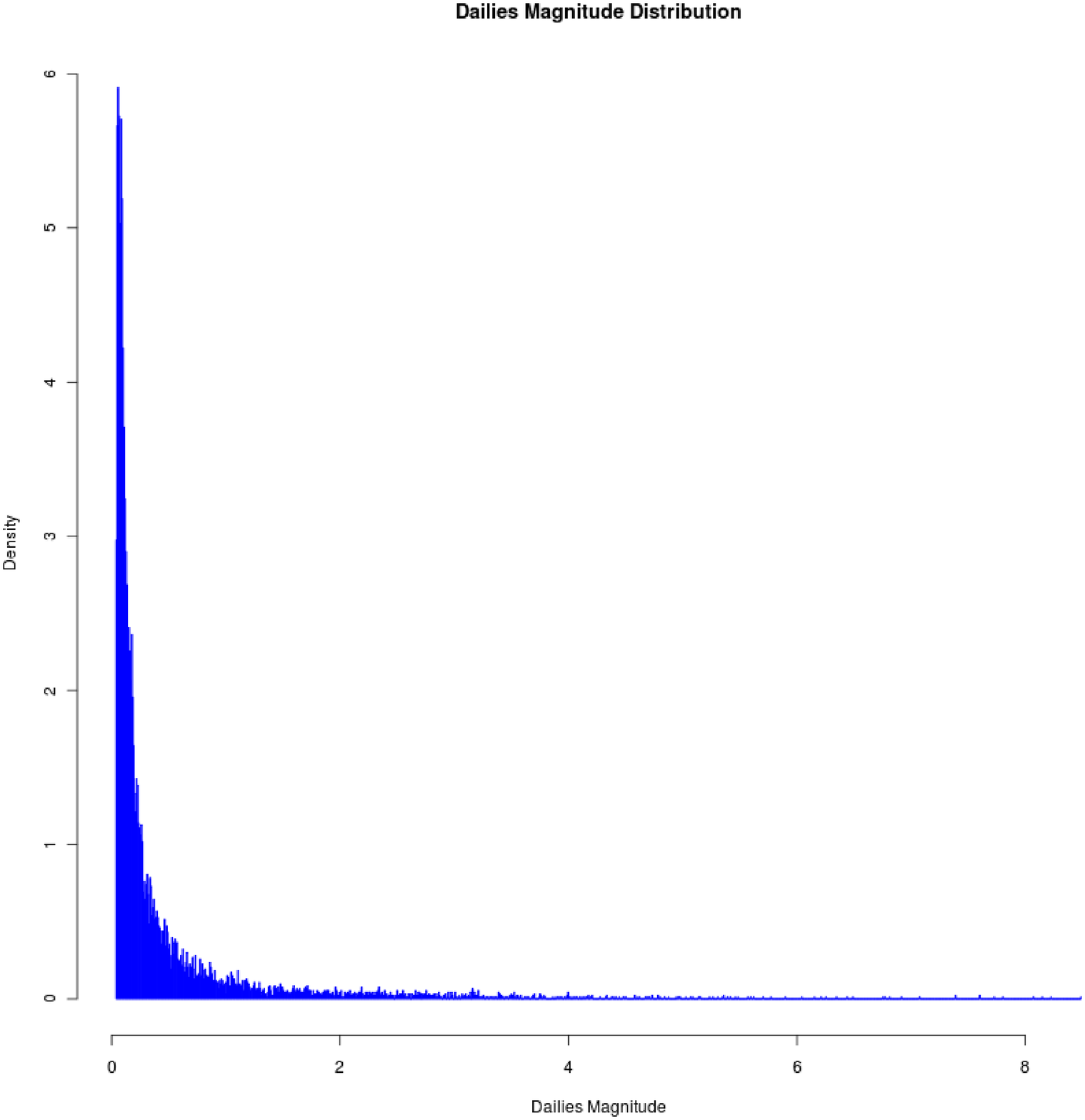}
\end{center}
\caption{Fame distribution of Dailies depository.}
\label{dailiesHist}
\end{figure}
}

\begin{list}{\labelitemi}{\leftmargin=1.5em} \itemsep -1pt

\item {\em Truncated Log-normal Distribution}:
Truncated log-normal distribution is a more general case. In Figure
\ref{hist-puerto}, the left tail could not go beyond 0 because fame
is always a non-negative number. Therefore, truncated log-normal
distribution \cite{arvid02} is introduced and in our case the
truncation point is 0. Moreover, truncated log-normal distribution
is more meaningful while the entity or group fame is small.
Formally, the probability density function of a left-truncated
log-normal distribution with truncation point $X_L$ is given by:
\begin{equation}
f_{LTN}(x)=  \left\{
    \begin{array}{ccc}
0, \ \ \ \ \ \ \ \ \ \ \ \ \ \ -\infty \leq x \leq x_L\\
\displaystyle \frac{f(x)}{\int_{x_L}^\infty \! f(x) \, dx}, \ \ \
x_L \leq x \leq \infty
\end{array}
    \right.
 \label{truncatedForm}
\end{equation}
where $f(x)$ is the probability density function of regular normal
distributions.

\end{list}

For entities with occasional mentions (like ``National Park Bank"),
we could use poisson distributions or power-law distributions to
approximate them. However, this category is less concerned by our
paper because people usually pay attention to important news
entities only.

In fact, log-normal and power-law distributions are discovered in
many physical, biological, economic and social systems. Most of our
distribution problems in this paper could also be answered by these
two popular distributions. Explanation of log-normal and power-law
distributions in social science could be found from
\cite{Mitzenmacher_abrief} and \cite{Sun04}.

\subsection{Entity Fame Distribution within a Group}
\label{powerlaw-distribution}

Now let's suppose each entity $E$ has a certain fame-level. Within a
news group, a large amount of entities have small fame-levels, and
only very few entities have large fame-levels. The number of
corresponding entities will exponentially decrease with the
increasing of fame. This is called power-law property. For example,
Figure \ref{dailies-dist} shows the histogram plot of all entities
in Dailies depository, which clearly indicates a power-law
distribution of entity fame.

Here we explain a little bit for the formulation of power-law
distributions. Initially, let us assume there is only one single
entity in the group. At each step, a new entity need to be mentioned
by news sources. With probability $\alpha<1$, the new entity is
going to be indeed a {\em new} entity chosen uniformly at random
from outside. With probability $1-\alpha$, the new entity is going
to be actually an {\em old} entity. This model is often called {\em
preferential attachment model} (\cite{Mitzenmacher_abrief},
\cite{Barabasi99mean-fieldtheory}), in which new entities tend to
attach to popular entities. This also agrees ``rich-getting-richer"
law. Eventually, above process generates a power-law distribution,
and the probability density function is defined as:
\begin{equation}
\displaystyle N(x) = cx^{-\lambda} \label{power-law}
\end{equation}
In this formula, $c$ and $\lambda$ are constants, and exponent
$\lambda$ is usually a positive number.

\begin{list}{\labelitemi}{\leftmargin=1.5em} \itemsep -1pt
\item {\em Truncated Power-law Distribution}: There is a tricky
problem for news group in terms of group definition. For example,
group {\em All U.S. Cities} perfectly follows our above preferential
attachment model and thus these cities' fame follows power-law
distribution. However, news group {\em Top 50 U.S. Cities} is
somewhat different because only some big cities are included in the
group. In this situation, a truncated power-law distribution
\cite{Mikael00} or a power-law tail should be applied to model this
group.

\end{list}

\fullversion{
\begin{figure}[pht]
\begin{center}
\includegraphics[height = 0.6\linewidth,width=0.9\linewidth]{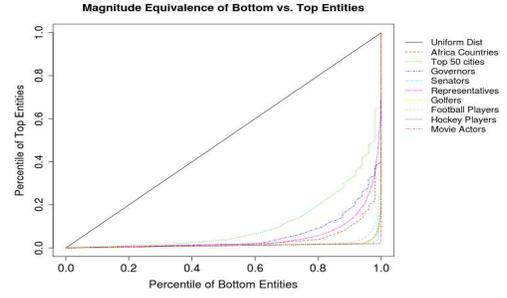}
\end{center}
\caption{Magnitude equivalence of bottom vs. top entities for
selective groups.} \label{allEquiv}
\end{figure}
}

\fullversion{
\begin{figure*}[ht]
\begin{minipage}[t]{0.32\linewidth}
\centering
\includegraphics[width=0.7\textwidth]{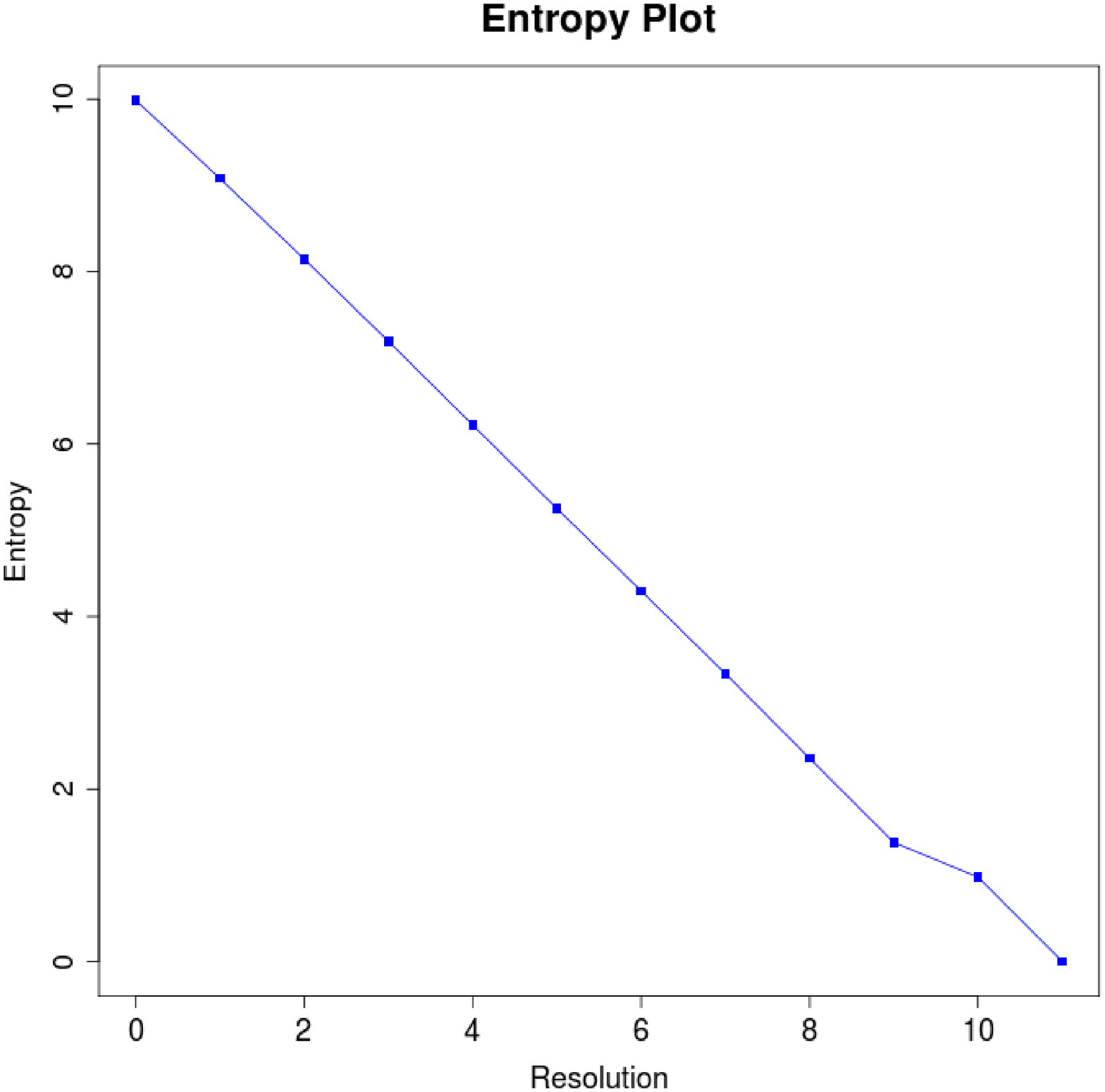}
\caption{Entropy plot for General Motors. The absolute value of
slope (information fractal dimension) is 0.92.\label{entropy-GM}}
\end{minipage}
\hfill
\begin{minipage}[t]{0.32\linewidth}
\centering
\includegraphics[width=0.8\textwidth]{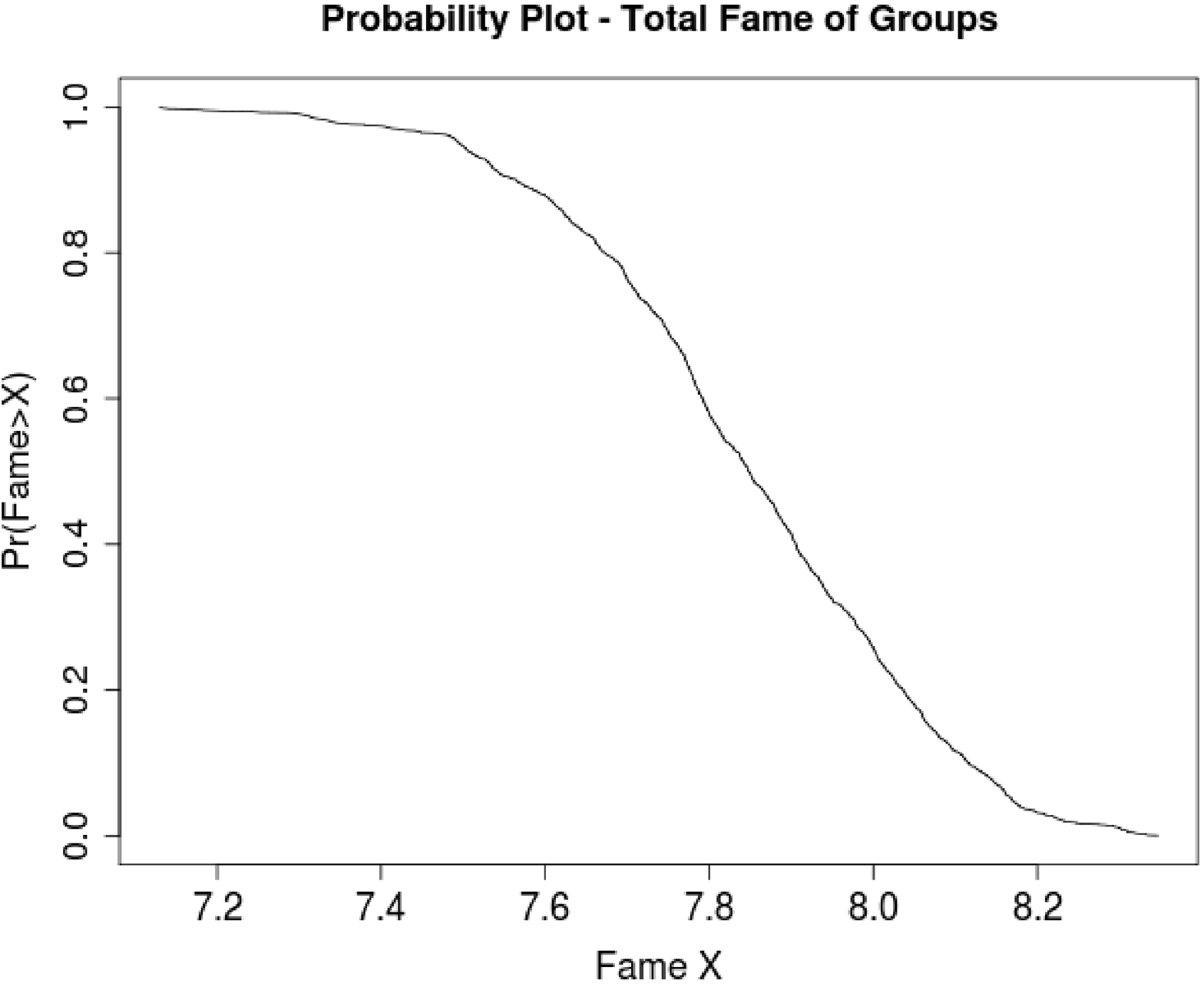}
\caption{Empirical curve of fame $X$ vs. the probability that the
total fame is greater than $X$ for group {\em Africa Countries}.
\label{plotFame_total_real_africa}}
\end{minipage}
\hfill
\begin{minipage}[t]{0.32\linewidth}
\centering
\includegraphics[width=0.8\textwidth]{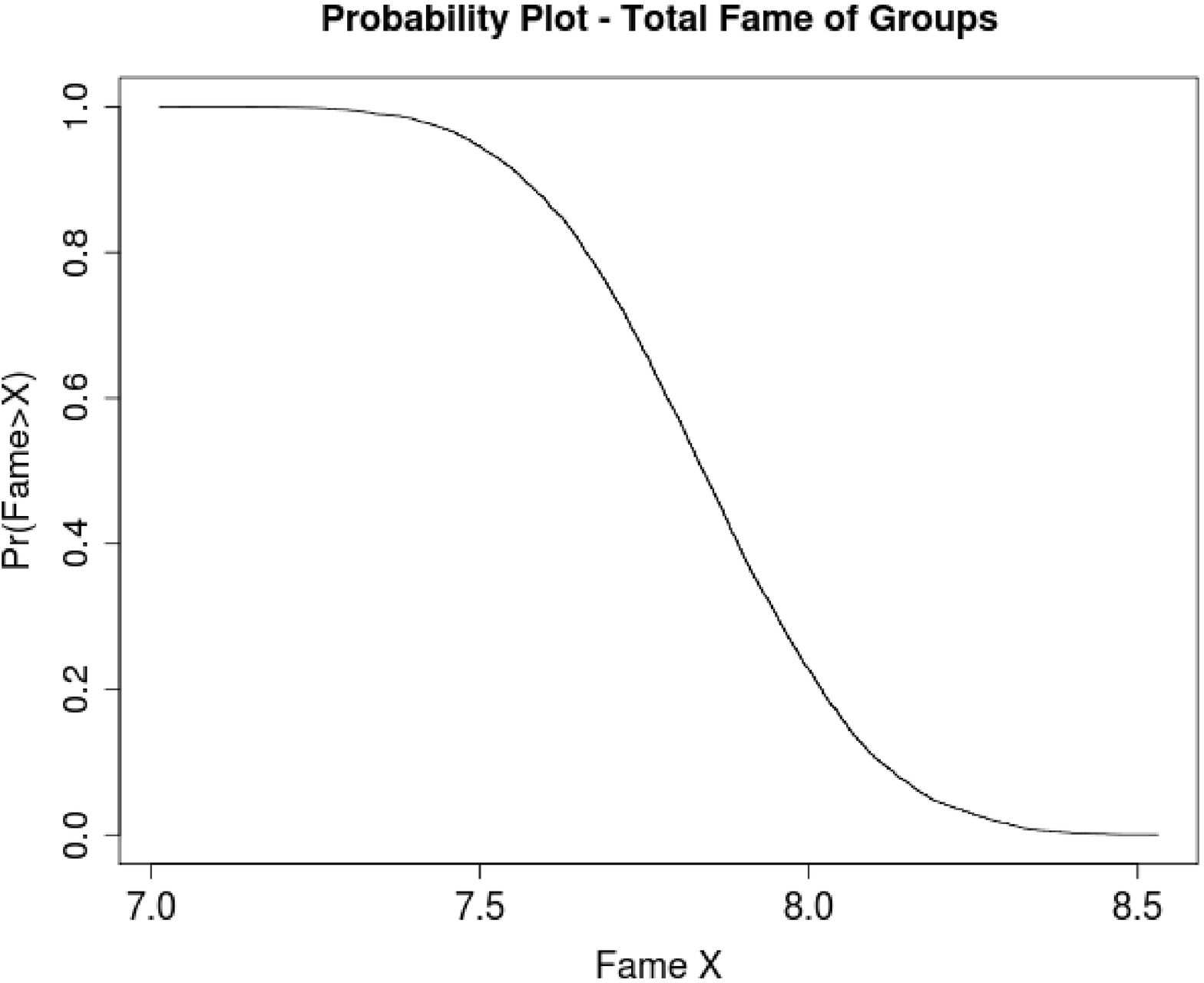}
\caption{Theoretical curve of fame $X$ vs. the probability that the
total fame is greater than $X$ for group {\em Africa Countries}.
\label{plotFame_total_theory_africa}}
\end{minipage}
\end{figure*}

}

The fame diversity within a group could be evaluated by the exponent
$\lambda$ of the histogram plot (or the slope in the corresponding
Log-Log plot). Another way to measure the inner group diversity is
to give bottom vs. top fame equivalence plots. Let us use
$F_{bottom}(G, \alpha\%)$ to denote the accumulated fame for the
bottom $\alpha\%$ entities, and use $F_{top}(G, \beta\%)$ to denote
the accumulated fame for the top $\beta\%$ for group $G$. If we make
$F_{bottom}(G, \alpha\%) = F_{top}(G, \beta\%)$, then we can plot a
line for all $(\alpha\%, \beta\%)$ pairs.

Figure \ref{allEquiv} shows the fame equivalence plots for 9
selective groups. We can know {\em movie actors} is the group with
the most significant fame diversity, while {\em Top 50 Cities} is
the group with the least fame diversity. In addition, governors,
senators, representatives are all politician groups, but governors
have little fame differences while senators have much bigger fame
differences.

\subsection {Group-Fame Probabilistic Modeling}

Now we will consider the fame of groups. For group $G$, we identify
three fame variables as below.

\begin{list}{\labelitemi}{\leftmargin=1.5em} \itemsep -1pt
\item {\em Total Group Fame $F_{G,T}$} - The total fame of all entities
in this group.

\item {\em Average Group Fame $F_{G,A}$} - The average fame for entities in
this group.

\item {\em Maximum Group Fame $F_{G,M}$} - The fame of the most famous
entity in this group.

\end{list}

\begin{figure}[htp]
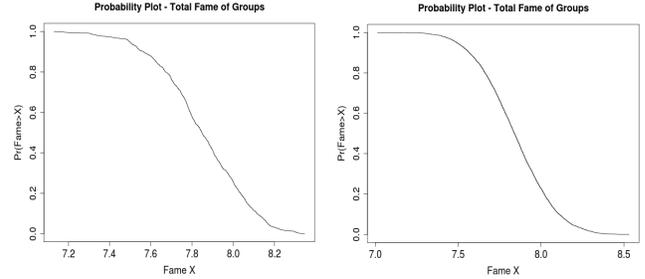

\centering
\includegraphics[width=0.49\linewidth,height=0.45\linewidth]{./figures/new/plotFame_total_real_africa.eps}
\hfill
\includegraphics[width=0.49\linewidth,height=0.45\linewidth]{./figures/new/plotFame_total_theory_africa.eps}
\caption{Empirical (left) and theoretical (right) curves of fame $X$
vs. the probability that the total fame is greater than $X$ for
group {\em Africa Countries}.} \label{plotFame_total_africa}
\end{figure}

\begin{figure*}[htp]
\centering
\includegraphics[height= 2.2in, width=0.9\linewidth]{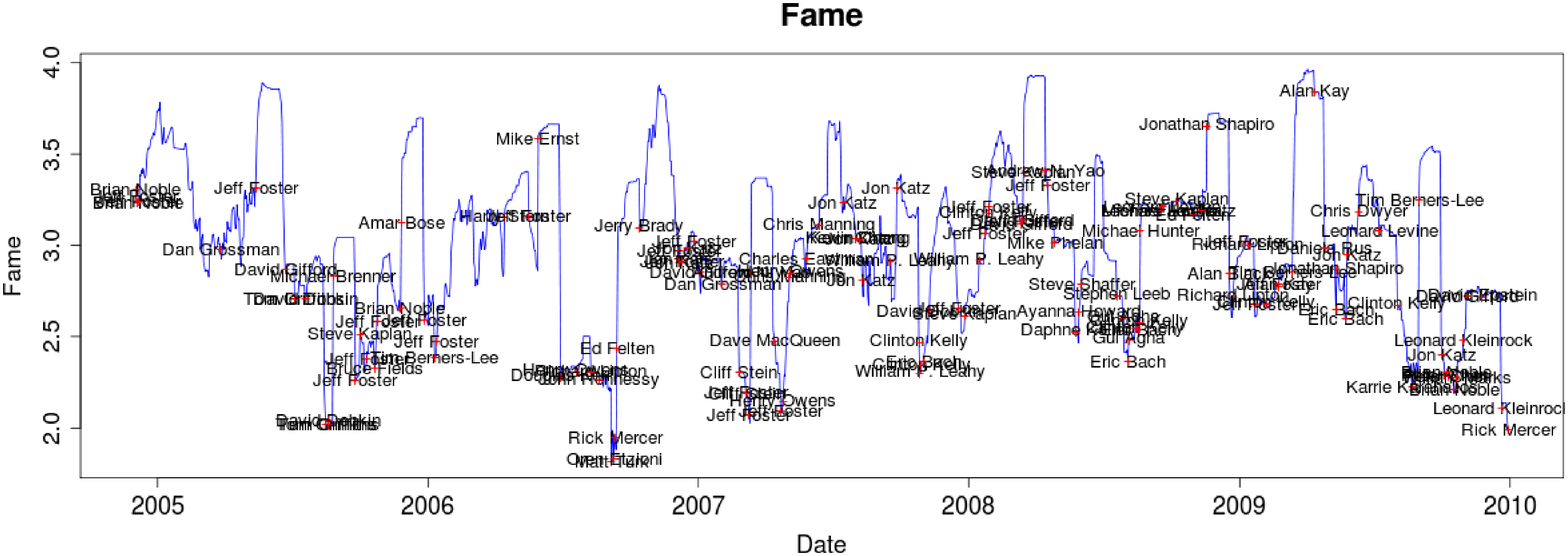}
\caption{The maximum fame time series of group {\em CS professors}.
The red markers indicate the start points from which the
corresponding entities have the biggest fame in this group, until
reaching the next red marker. } \label{fameMove}
\end{figure*}

An interesting question is, can we estimate the probability that the
total fame, average fame, or the maximum fame of a certain group is
greater than a fame level $X$? We denote the probability as
$Pr(F_{G,k}>X)$,where $k$ could be replaced with $T$, $A$, or $M$.
Clearly, we have $Pr(F_{G,k}>0)=1$ and $Pr(F_{G,k}>\infty)=0$.

According to Section \ref{distribution}, we know that the total,
average, and maximum fame of groups all approximately follow
log-normal distributions. Then using groups' training data, we can
get distribution $F_{G,k} \sim N(\mu, \sigma^{2})$. Assuming
${\Phi_F}(X)$ is the cumulative distribution function of $F_{G,k}$,
we have
\begin{equation}
Pr(F_{G,k}>X) = 1- {\Phi_F}(X) \label{fameProb}
\end{equation}
With applying this approach, we get the empirical and theoretical
probabilities of the total fame of {\em Africa Countries}, as shown
in Figure \ref{plotFame_total_africa}. The empirical curve is
calculated from the real news data, while the theoretical curve is
calculated from our log-normal model \ref{fameProb}. The two curves
are pretty similar, and thus the log-normality of group fame
distributions could be validated.

\fullversion{ If we compare {\em Africa Countries} with all other
type of news groups, we will get the empirical and theoretical
probability plots shown in Figure \ref{plotFame_total_africa}. From
the plots, we can see that the total fame of 439 representatives is
similar to that of 49 governors, but the total fame of 105 senators
is much higher. We also know governors have higher fame than
representatives in average. Among all the 9 groups, movie actors
have the biggest total fame than all other groups, which is
reasonable because the population of movie actors is much larger
than other groups. }

\subsection {Fame Change Over Time} \label{fameChange}

Another important problem we study is fame movement over long time.
For example, Figure \ref{fameMove} shows the maximum fame time
series of group {\em CS professors} from 2005 to 2009, with fame
window size 1 month. We should notice that the most common names are
filtered out. For example, ``Michael Jordan" may refer to either a
basketball players or a CS professor, so these kind of names are not
counted in.

\begin{figure}[htp]
\centering
\includegraphics[width=1.1\linewidth]{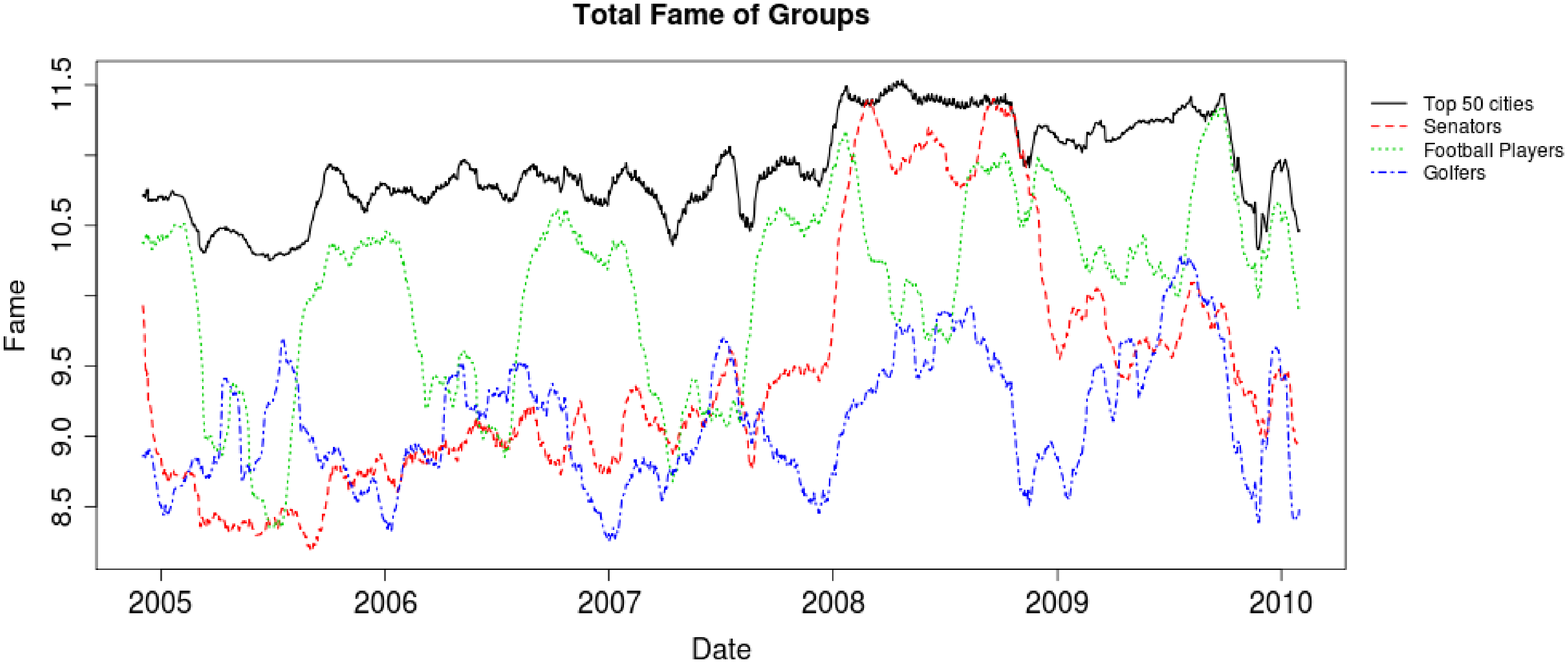}
\caption{The group total fame over 5 years for 4 different groups.
The fame window size is 1 month.} \label{groupTotalFame}
\end{figure}

An interesting idea is to select some groups and compare their group
total, maximum, and average fame. Figure \ref{groupTotalFame} is the
example of total fame. We can see below highlights:

\begin{list}{\labelitemi}{\leftmargin=1.5em} \itemsep -1pt
\item The movement of group {\em Top 50 cities} is more smooth than
other groups.
\item For senators, there is a big and durable jump in 2008 in Figure \ref{groupTotalFame}. This is
because Barack Obama was ever a senator and he was running for the
2008 presidential election at that time.
\item Sportsman groups are even more interesting. Figure
\ref{groupTotalFame} shows the fame of {\em Football Players} and
{\em Golfers} fluctuates periodically. Basically the periodicity is
caused by sport seasons. For example, the National Football League
 season is usually from September to the next January, which matches
the green line in Figure \ref{groupTotalFame}.
\end{list}

Indeed, the fame fluctuation over time is very interesting, and with
which we could figure out some significant events and lots of other
useful information of news.

\section{HMM-based News Generation Model and Group Maximum Fame Forecasting} \label{hmmBased}

Although we can use log-normal approaches to describe news frequency
or fame distributions, we still need to develop a practical news
generation model because the simple log-normal model has several
critical problems. For example, the log-normal model cannot simulate
the bunchy arrivals of news peaks, and it cannot imitate the trend
of news start, reaching peak, and decay curve. Actually log-normal
model is equivalent to a geometric Brownian motion, which is not
exactly true for news generation.

Here we will propose an innovative Hidden Markov Model (HMM) to
model news generations. The idea of the HMM model is that typically
a news entity should be in one of two possible states, {\em normal}
state or {\em peak} state.  In this section, first we will provide
pulse detection algorithm and pulse curve fitting algorithm, then we
will describe our HMM model in detail. \fullversion{The two
algorithms will be used in the training phase of the HMM model, so
essentially they are part of our HMM model.}

\subsection{Pulse Detecting and Fitting} \label{pulse}

There are numerous methods to detect pulses from time series. But
here we just propose a very straightforward approach to identify
pulses. The detailed algorithm is shown in Algorithm \ref{alg0}.

\begin{algorithm}[h!]
\caption{Pulse detection algorithm}
\algsetup{linenosize=\footnotesize} \footnotesize \label{alg0}
\begin{algorithmic}[1]
\scriptsize

\REQUIRE Parameters: multiple $K$ for multiplying time series
standard deviation, number $t$ for neighbor grouping, and length $N$
for moving average calculation.

\STATE {\bf Peak detection}: Identify local maximums in time series,
which should be larger than 1) $K$ times of the time series standard
deviation ($K\sigma$), and 2) neighbor data points of both sides.
$K$ is a selected integer.

\STATE {\bf Peak grouping}: Adjacent peaks within backward or
forward $t$-distance should be grouped together to form a single
peak group. Here $t$ is a user specified integer.

\STATE {\bf Pulse identification}: Compute the backward length-$N$
moving average from the first data point of the peak group, and the
forward length-$N$ moving average from the last data point of the
peak group, absorb neighbor data points until the moving average
starts to increase. $N$ is also an elaborately selected integer.
Now, a completed pulse could be identified.

\RETURN Detected pulses.

\end{algorithmic}
\end{algorithm}

\fullversion{
After pulses are identified, we will be able to
analyze some pulse-relevant distributions:

\begin{enumerate}
\item Peak interval distribution: The distribution of distance for two
neighboring peak points.
\item Pulse size distribution: Pulse size is defined as:
\begin{equation}
\displaystyle S=\sum_{i=1}^m{f_i}
\end{equation}
$f_1, f_2,...,f_m$ are data points to form the completed pulse.
\item Pulse height distribution: The distribution of the maximum
reference count within pulses.
\end{enumerate}

Based on the analysis of Section \ref{distribution}, peak interval
should follow power-law distributions, while pulse size and pulse
height will follow log-normal distributions. Indeed, our
experimental results prove these results as well.
}

\begin{figure}[pht]
\begin{center}
\includegraphics[height = 1.3in, width=0.65\linewidth]{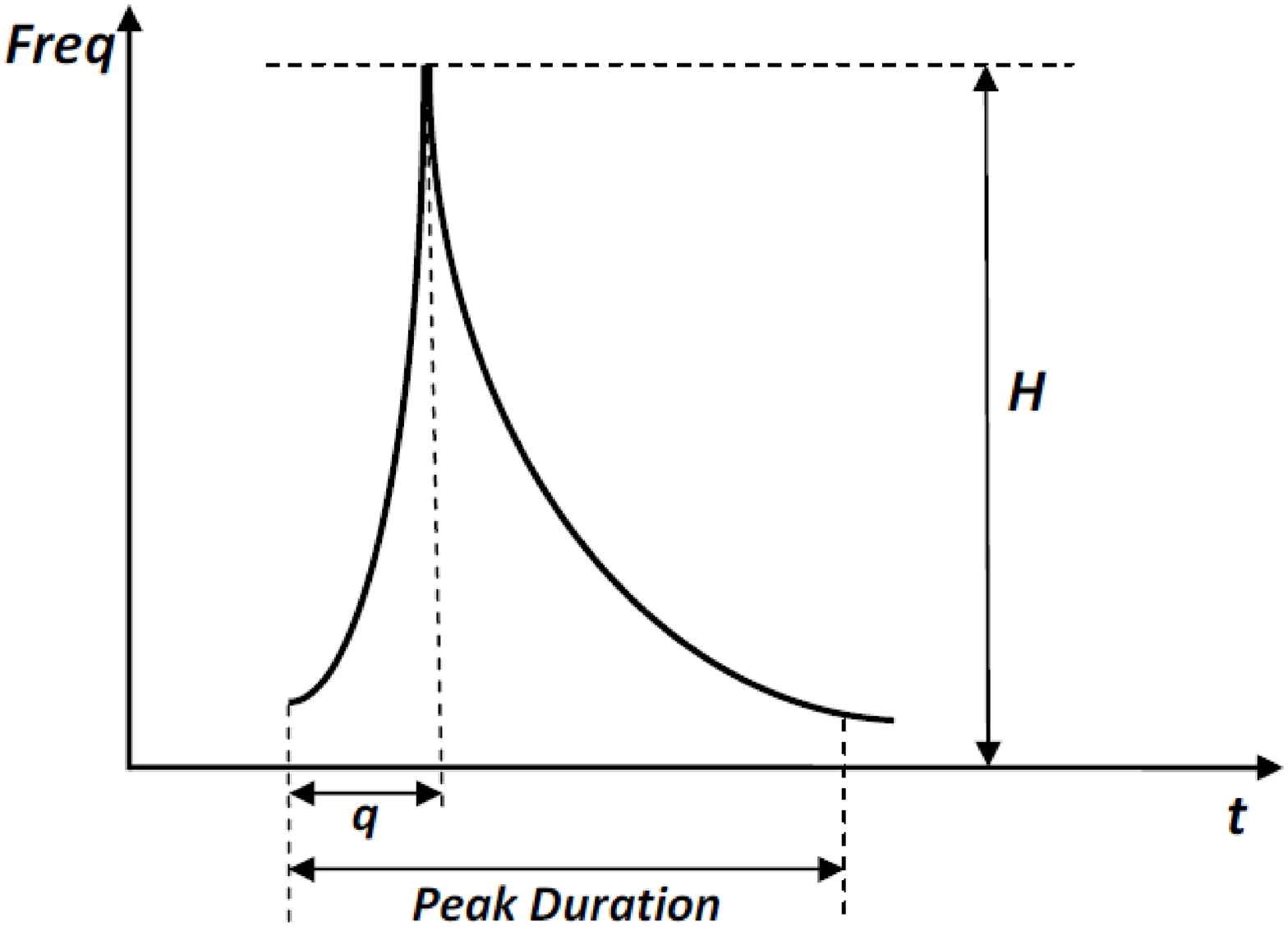}
\end{center}
\caption{A graph to illustrate the structure of pulses. Here $q$ is
the number of days to reach the peak point, and $\mathcal {H}$ is
the height of the peak.} \label{peak}
\end{figure}

\begin{figure}[ht]
\begin{center}
\includegraphics[height=1.5in, width=1.04\linewidth]{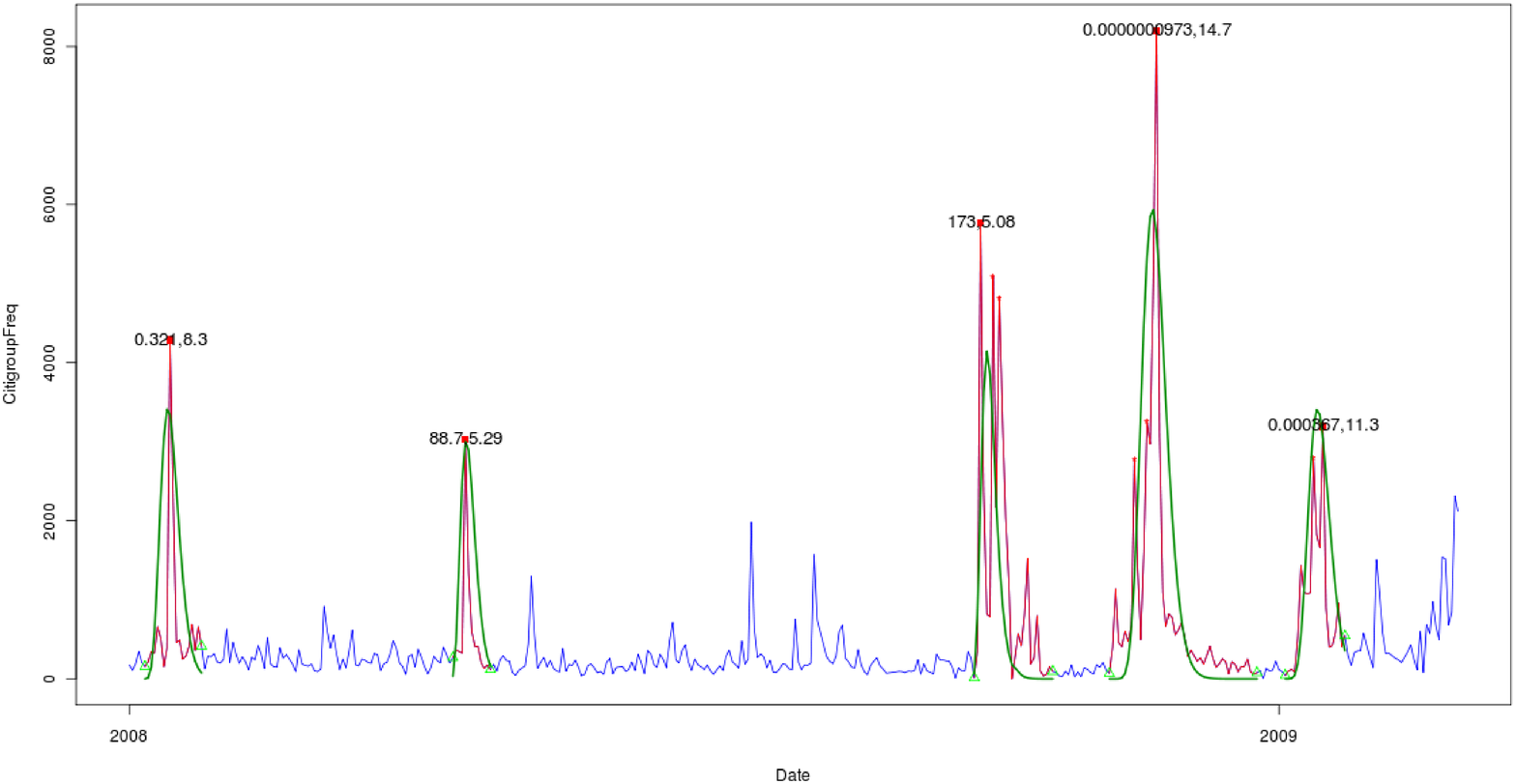}
\end{center}
\caption{Pulses identified from Citigroup time series. The red
sections in the time series indicate the pulses identified by our
algorithm, and the green curves are fitted by our fitting
methodology. } \label{citi-pulseFitting}
\end{figure}

A further question is that how to use function to fit identified
news pulses. Leskovec et al. (\cite{Leskovec09}) proposed a
mathematical model to imitate the procedure of news threads' start,
peak, and decay. They argue that two minimal ingredients should be
taken into account to simulate news cycles, imitation effect and
recency effect. Imitation effect means that different news sources
imitate one another, and recency effect means that newer threads are
favored to older ones. If we define a monotonically increasing
function $f(\cdot )$ and a monotonically decreasing function
$\delta(\cdot )$ to mimic the two ingredients separately, we can
derive below equation:
\begin{equation}
\displaystyle x(t+1) = cf(x(t))\delta(t)
\end{equation}
Here $x(t)$ is the news reference for time $t$, $x(t+1)$ is the news
reference for time $t+1$, and $c$ is a normalizing constant. Thus a
differential equation could be derived: $\displaystyle \frac{dx}{dt}
= cf(x)\delta(t)-x$. With assuming $f(x) = qx$ and $\delta (t) =
t^{-1}$, then we get:
\begin{equation}
x=At^q e^{-t} \label{pulseFitting}
\end{equation}
The function reaches peak while $t=q$, and the peak value is
$\displaystyle \mathcal {H}=x(q)=A(\frac{q}{e})^q$, which is also
the height of peak. Therefore, we get $\displaystyle A=\mathcal
{H}(\frac{e}{q})^q$, and thus Equation \ref{pulseFitting} becomes
\begin{equation}
\displaystyle x(t)=\mathcal {H}(\frac{e}{q})^qt^qe^{-t}=\mathcal
{H}(\frac{et}{q})^qe^{-t} =\mathcal
{H}(\frac{t}{q})^qe^{q-t}\label{pulseFitting_new}
\end{equation}
In this model, we need to fit two parameters, $q$ and $\mathcal
{H}$, in which $q$ is the days to reach peak and $\mathcal {H}$ is
the height of the peak. Formula \ref{pulseFitting_new} is a little
bit more complicate than Formula \ref{pulseFitting}, but it is
practically more meaningful because we can use historical data to
estimate the distribution of $q$ and $\mathcal {H}$ for a certain
magnitude of entities. News reference monotonically increases before
day $q$ but decreases after day $q$, just like the pulse structure
shown at Figure \ref{peak}.

Figure \ref{citi-pulseFitting} is an empirical example to show how
the pulses are identified, and how well pulses fit to the Formula
\ref{pulseFitting}. The green curves are the fitted value of Formula
\ref{pulseFitting}, which are calculated by the least-square
non-linear regression methods. The three parameters we used for
pulse detection are $K=5$, $t=20$, and $N=10$ respectively. If we
track down the five pulses, we will see a story chain of Citigroup
in 2008.

\fullversion{
\begin{itemize}
\small
\item Citigroup suffers nearly \$10B billion loss, which is worst ever - Jan. 15, 2008

\item Citigroup posts another loss, cuts 9000 more jobs - Apr. 18, 2008

\item Citigroup to Buy Wachovia¡¯s Bank Assets for \$1 a Share -
Sept. 29, 2008

\item The U.S. government announced a massive bailout of Citigroup
- Nov. 24, 2008

\item Citigroup loses \$8.3 billion, moves to break up - Jan. 16, 2009
\end{itemize}
}

\subsection{Hidden Markov Model of News} \label{HMMnews}

\begin{figure}[pht]
\begin{center}
\includegraphics[height=0.9in,width=0.7\linewidth]{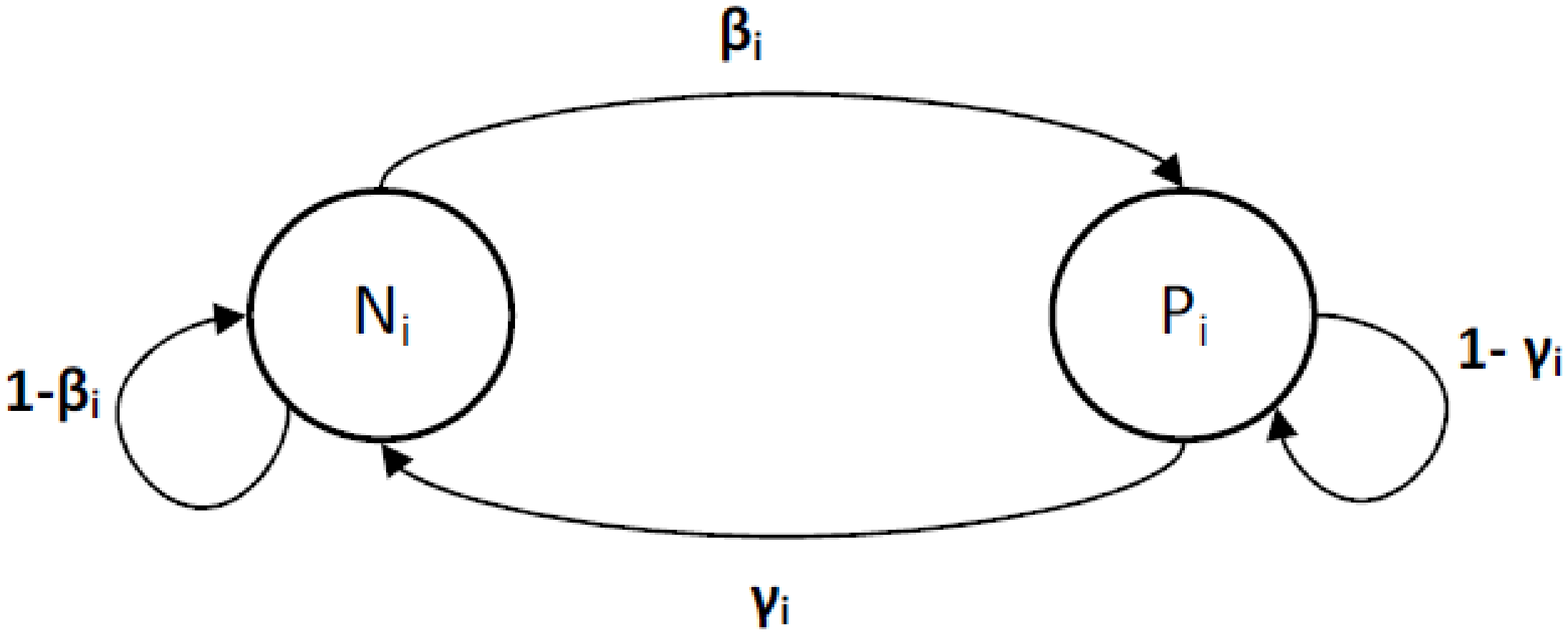}
\end{center}
\caption{State transitions of our HMM model. $N_i$ is the normal
state, and $P_i$ is the peak state.} \label{microStates}
\end{figure}

Now we are ready to design news HMM model. The detail of our HMM
model is shown in Figure \ref{microStates}. There are two states
$N_i$ and $P_i$ to denote the normal state and peak state
respectively. Usually an entity is in the normal state $N_i$, but it
will jump to $P_i$ state while big pulses are generated.  The
transition probabilities are defined by matrix
\begin{equation}
\Gamma=\left( \begin{array}{ccc}
1-\beta & \beta\\
\gamma & 1-\gamma \end{array} \right) \label{transition}
\end{equation}
Here $\beta$ is the transition probability from state $N_i$ to state
$P_i$. Usually $\beta$ is very small because entities are not very
exciting in most of the time. $\gamma$ is the transition probability
from state $P_i$ to state $N_i$, and it should be a probability
close to 1 because entities always tend to calm down within a
shorter or longer time after a pulse.

\begin{figure}[pht]
\begin{center}
\includegraphics[width=0.85\linewidth]{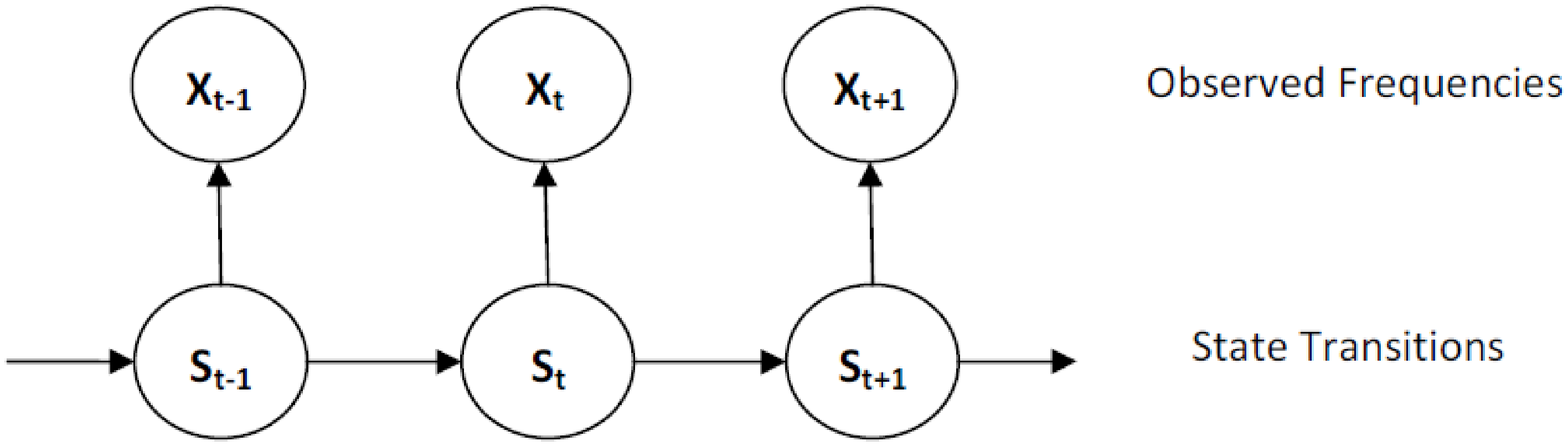}
\end{center}
\caption{The hidden Markov chain in $HMM$. $S_j$ chain is state
series, which could be either $N_i$ or $P_i$. $X_j$ chain is the
observed frequency/fame series. } \label{hmm}
\end{figure}

The hidden Markov model could be denoted as $\{X_t: t\in
\mathbb{N}\}$. The model consists of two parts: firstly an
unobserved `state process' $\{S_t:t=1,2,...\}$ satisfying the Markov
property, and secondly the `state-dependent process' $\{X_t:
t=1,2,...\}$ such that, when $S_t$ is known, the estimation of $X_t$
depends only on the current state $S_t$ and not on any previous
states or observations. The $HMM$ chain to model state transitions
is shown in Figure \ref{hmm}, in which the states $S_j$ could be
either $N_i$ or $P_i$.

\begin{figure*}[htp]
\centering
\subfigure[f(G,0,20,1,1)] 
{
    \label{fig:a}
    \includegraphics[height=0.20\linewidth,width=0.23\linewidth]{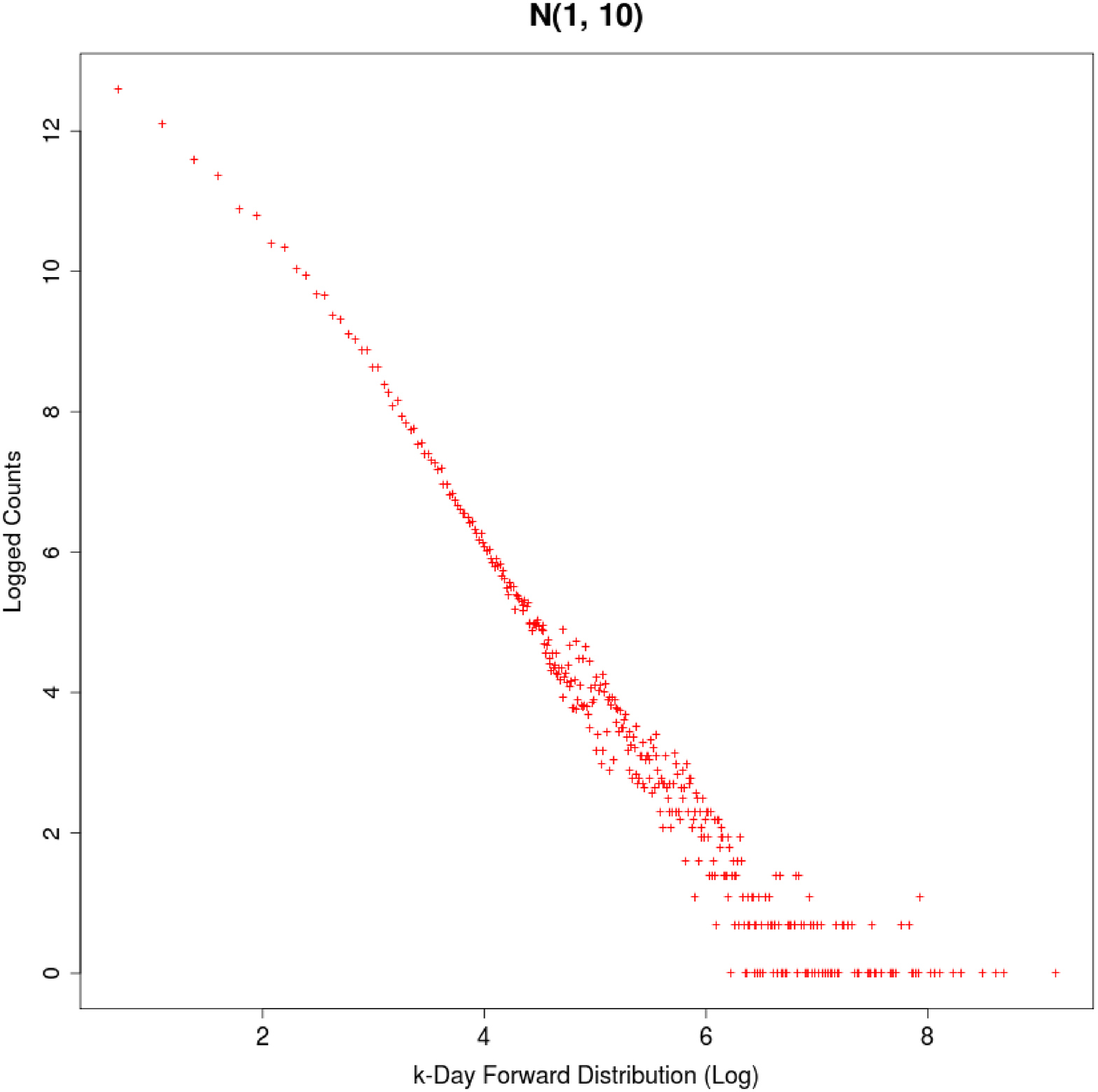}
} \hspace{-0.3cm}
\subfigure[f(G,50,100,1,1)] 
{
    \label{fig:b}
    \includegraphics[height=0.20\linewidth,width=0.23\linewidth]{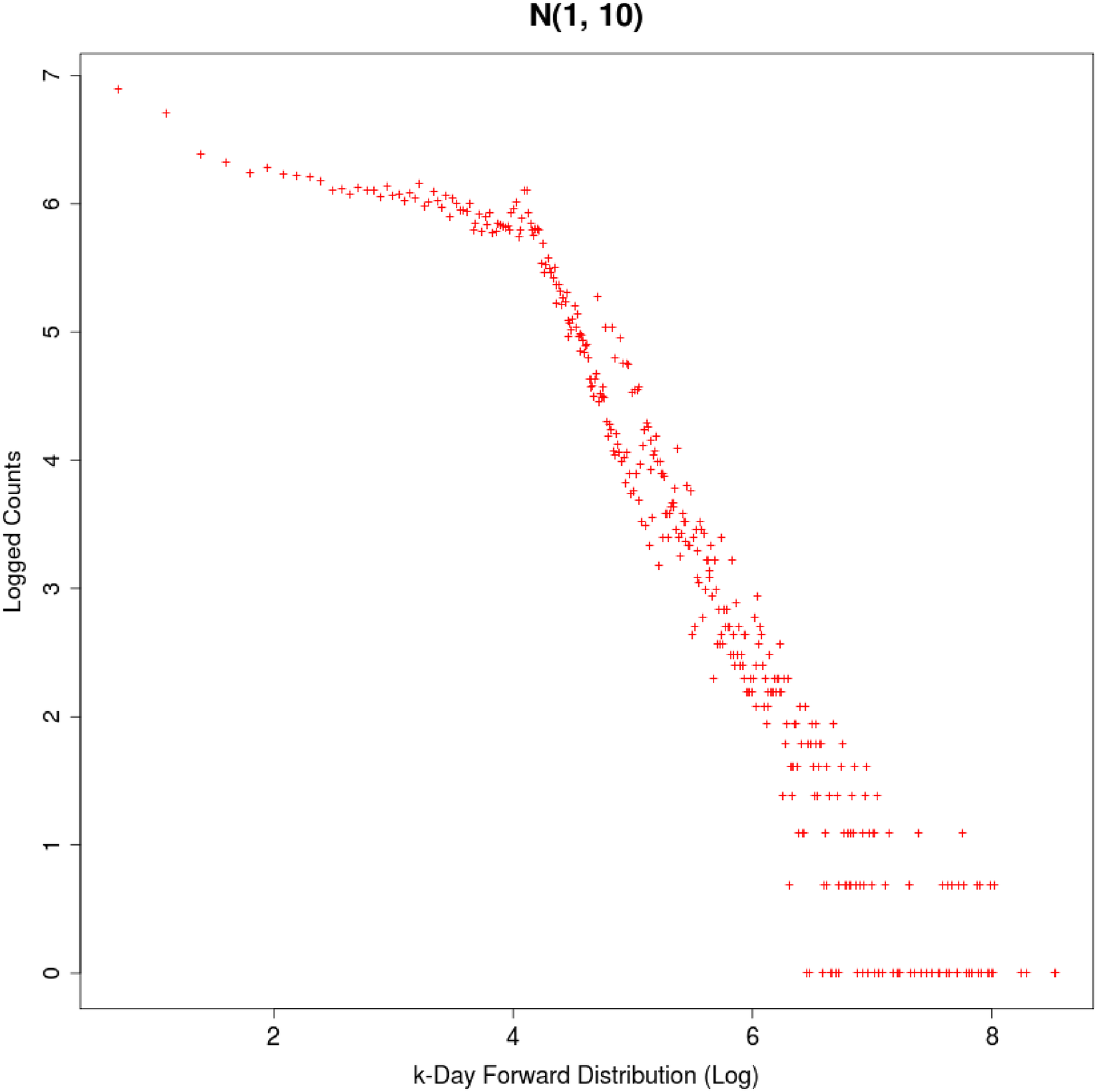}
} \hspace{-0.3cm}
\subfigure[f(G,30,30,1,1)] 
{
    \label{fig:c}
    \includegraphics[height=0.20\linewidth,width=0.23\linewidth]{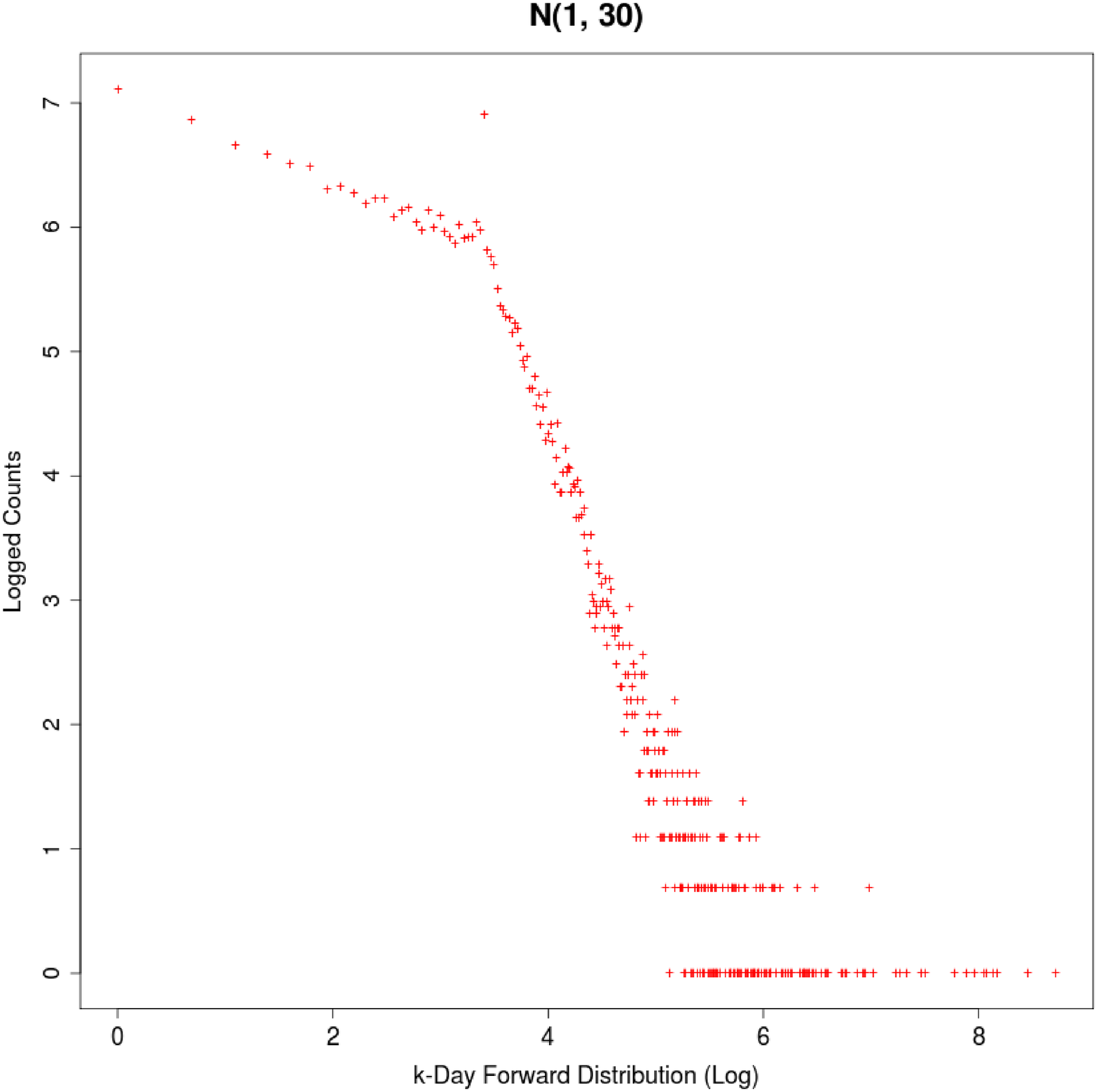}
} \hspace{-0.3cm}
\subfigure[f(G,20,50,60,5)] 
{
    \label{fig:d}
    \includegraphics[height=0.20\linewidth,width=0.23\linewidth]{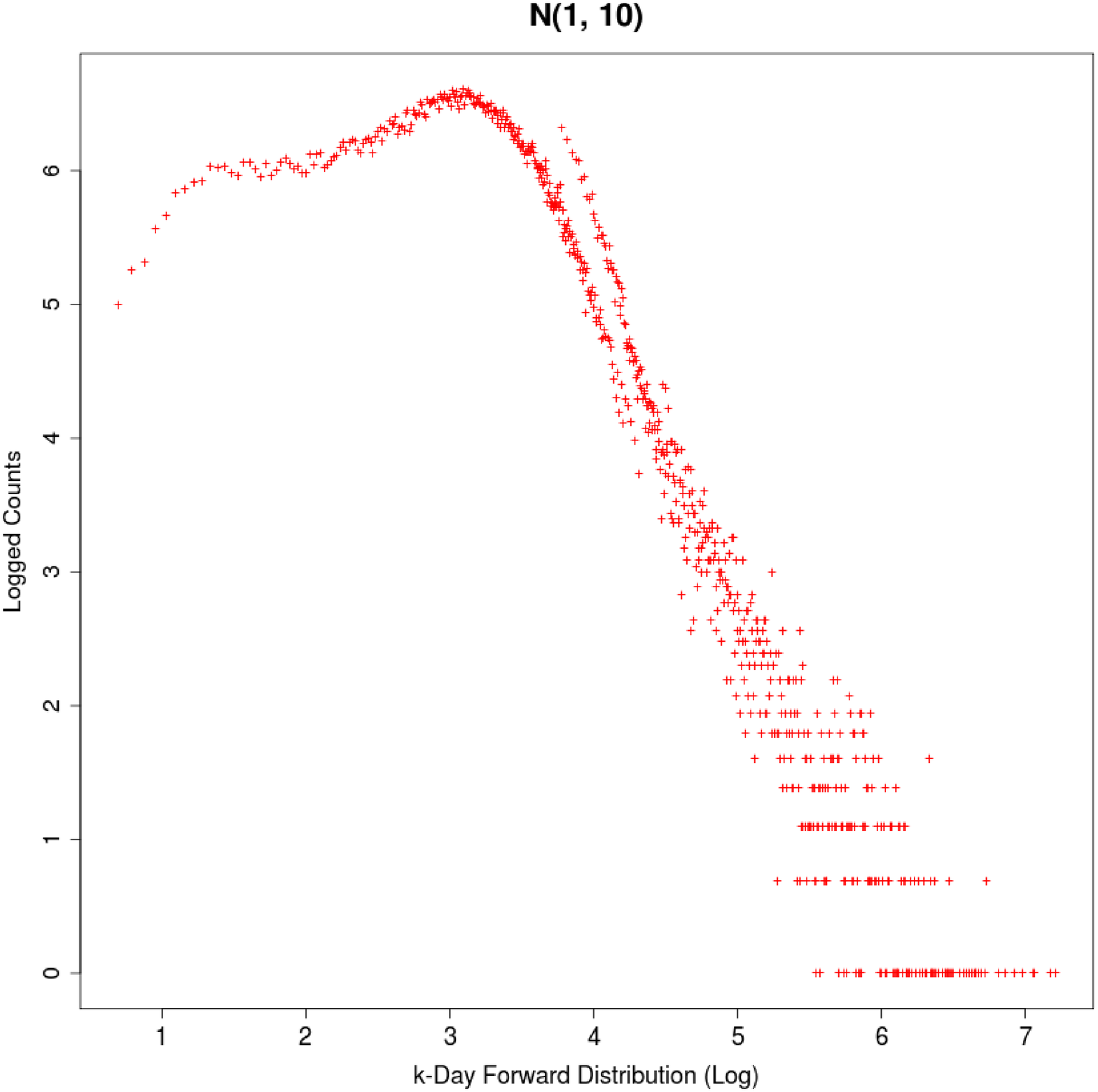}
} \\
\subfigure[N(G,0,20,1,1)] 
{
    \label{fig:e}
    \includegraphics[height=0.21\linewidth,width=0.24\linewidth]{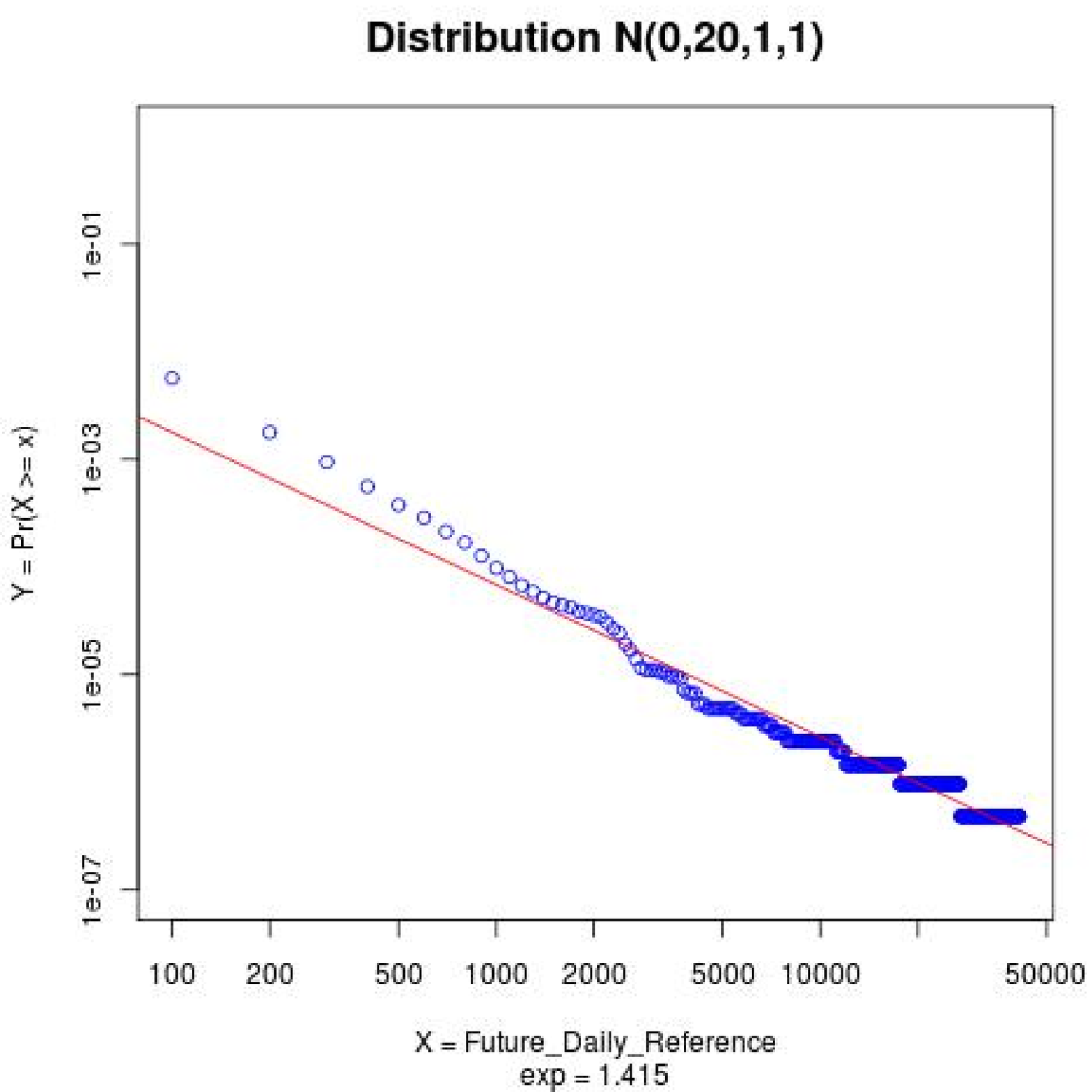}
} \hspace{-0.3cm}
\subfigure[N(G,50,100,1,1)] 
{
    \label{fig:f}
    \includegraphics[height=0.21\linewidth,width=0.24\linewidth]{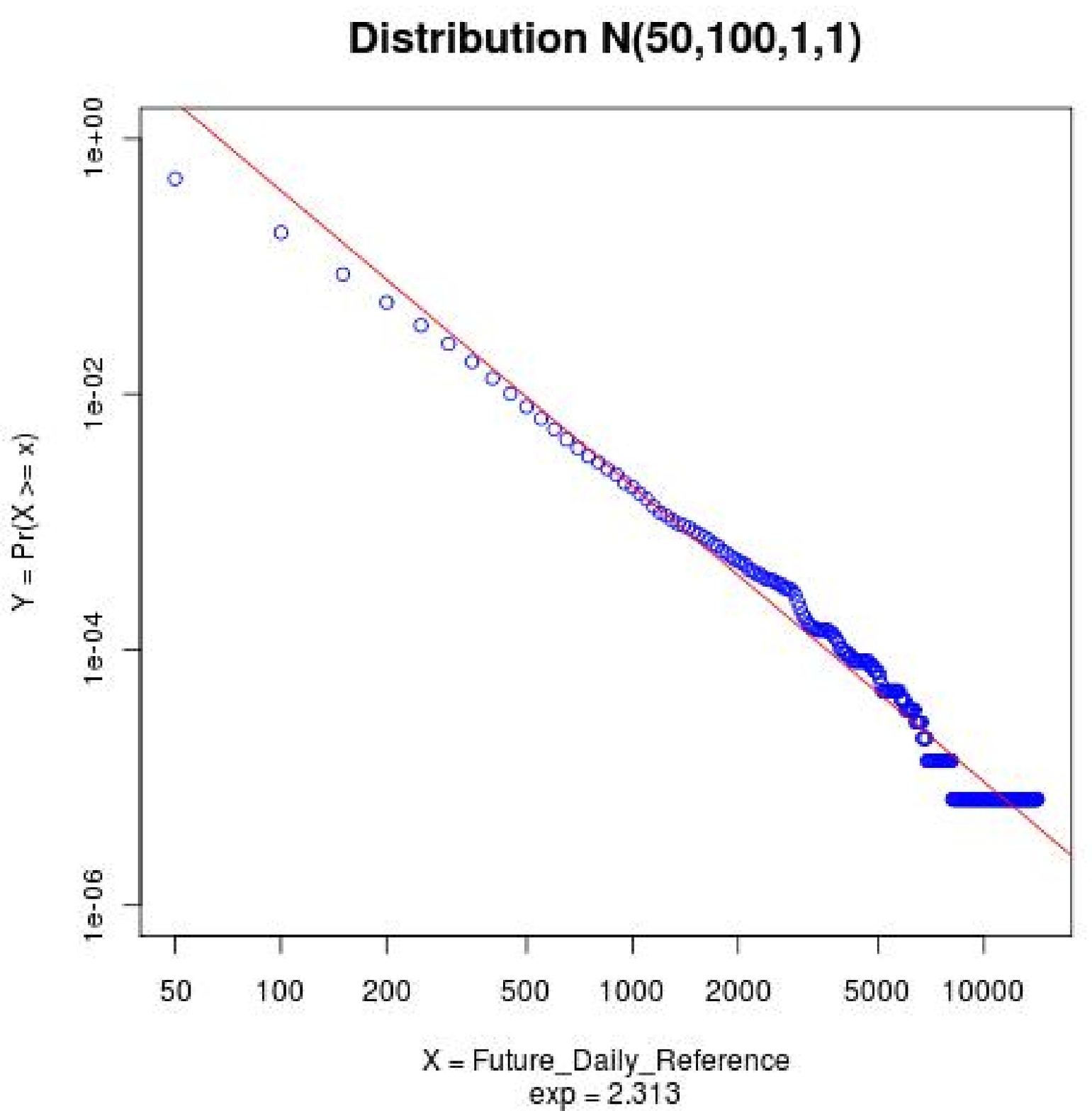}
} \hspace{-0.3cm}
\subfigure[N(G,30,30,1,1)] 
{
    \label{fig:g}
    \includegraphics[height=0.21\linewidth,width=0.24\linewidth]{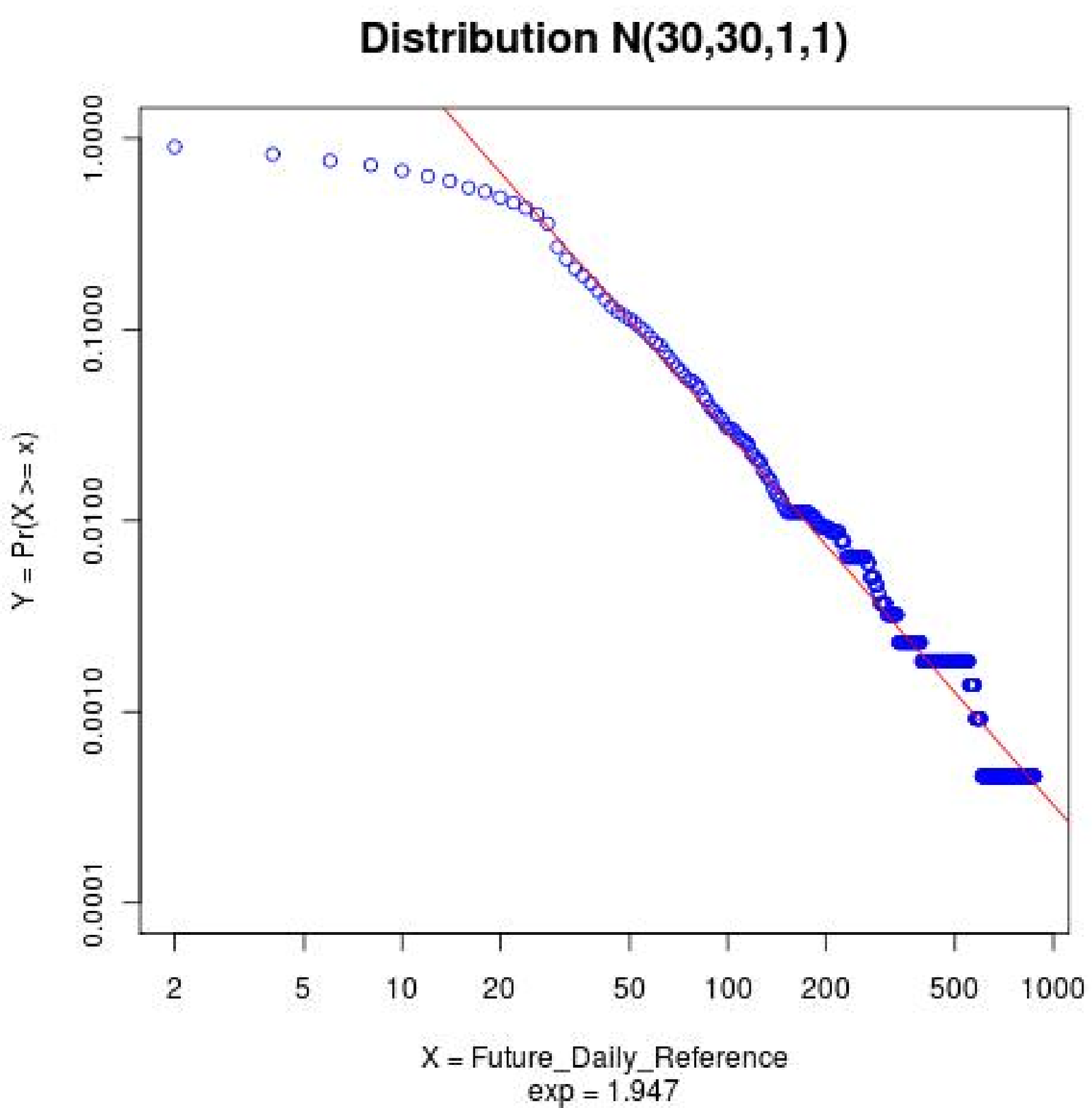}
} \hspace{-0.3cm}
\subfigure[N(G,20,50,60,5)] 
{
    \label{fig:h}
    \includegraphics[height=0.21\linewidth,width=0.24\linewidth]{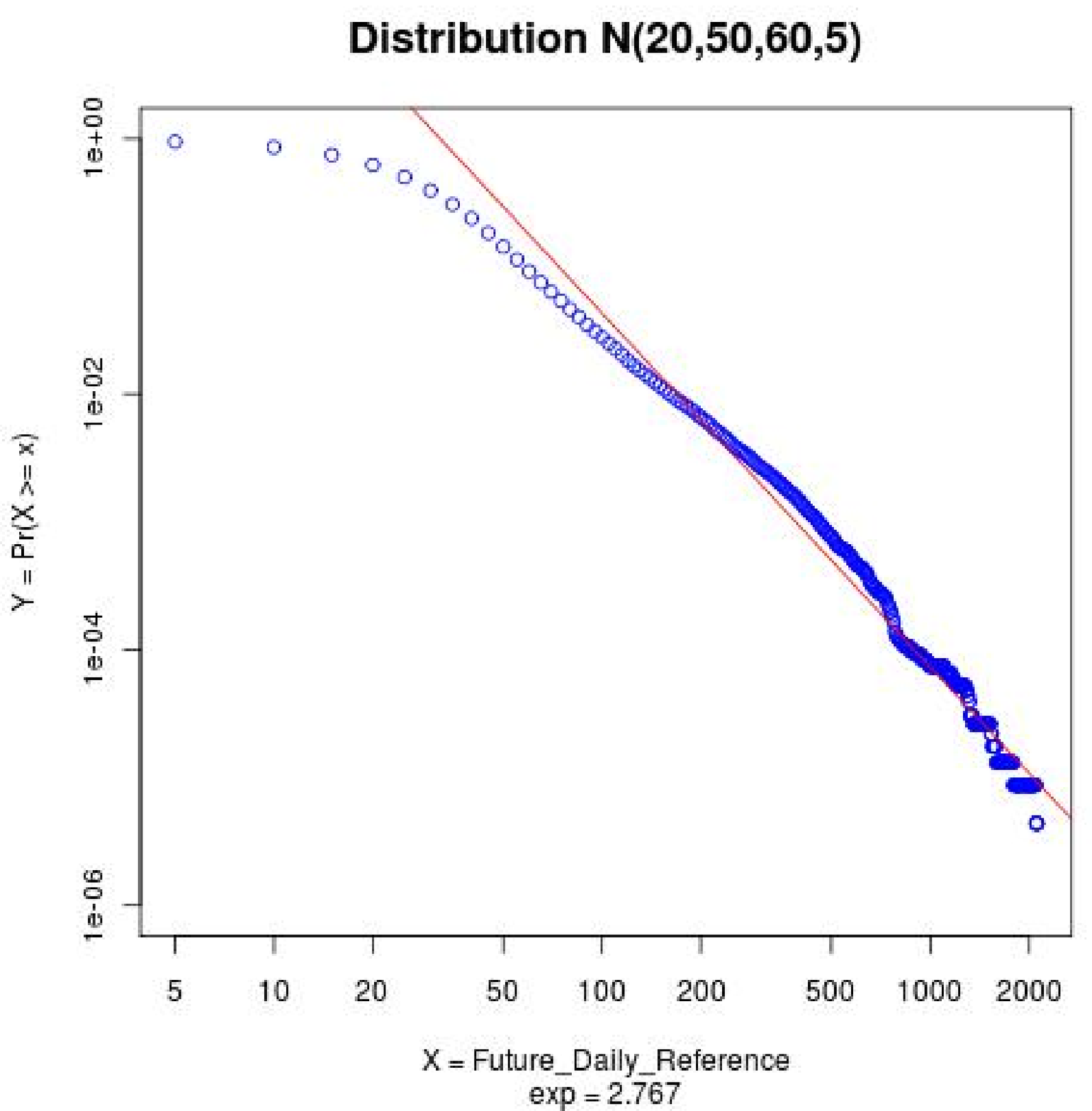}
}\caption{The future frequency distribution for entities in our
Dailies depository. The log-log plots show that the histogram plots
of $f(G, m_l, m_u, w_m, w_f)$ have power-law tails (Subfigure
\ref{fig:a} to \ref{fig:d}). Moreover, distributions $N(G, m_l, m_u,
w_m, w_f)=Pr(X>x)$ also have power-law tails (Subfigure \ref{fig:e}
to \ref{fig:h}).}
\label{forwardDist} 
\end{figure*}

\fullversion{
{\bf Theorem 1} {\em If the $HMM_i$ starts from state
$N_i$, the expected time until reaching state $P_i$ is
$\displaystyle \frac{1}{\beta}$}.

{\bf Proof 1} We use a recursion approach to solve this problem.
Let's assume the expected time to reach $P_i$ is $\nu$. Starting
from $N_i$, if the process moves to state $P_i$, then we have done
and $\nu=1$. Otherwise, if the process stays in state $N_i$, we are
in the same situation as the very beginning, say, we still need time
$\nu$ to reach state $P_i$. Therefore, we have the equation

\begin{equation}
\nu = 1+\beta*0+(1-\beta)*\nu
\end{equation}
Thus the expected time until reaching state $P_i$ is $\displaystyle
\nu=\frac{1}{\beta}$.

}

In our model, we pay much more attention to peak states than normal
states, because peaks are the most important parts of an entity's
time series. Therefore, we just use a log-normal distribution or
Geometric Brownian Motion to approximate the fluctuation of news
while it is in the normal state. In the peak state $P_i$, we use
simulated parameters described in Subsection \ref{pulse} to generate
news pulses.

\fullversion{
\begin{figure}[htp]
\centering
\includegraphics[height= 2.3in, width=0.8\linewidth]{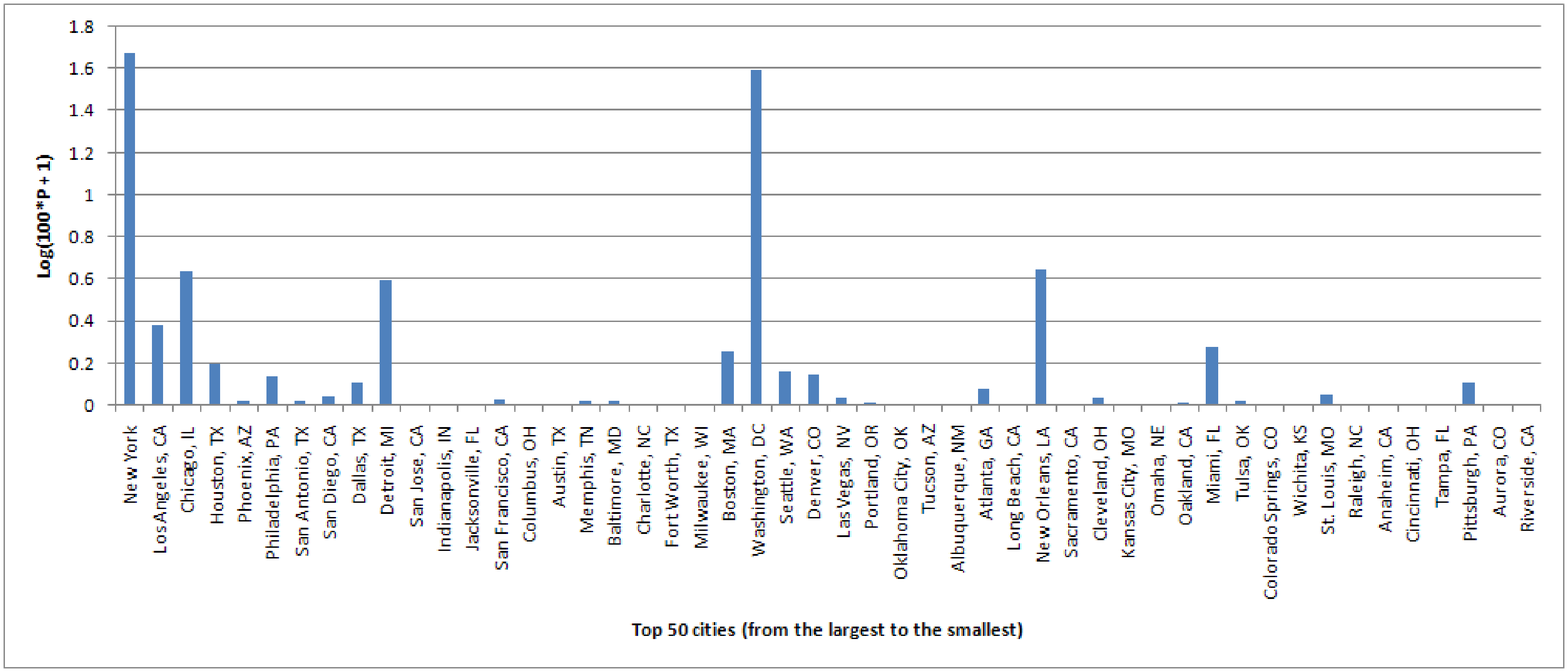}
\caption{Top 50 Cities' population vs. their fame in news.}
\label{cityPopu}
\end{figure}
}

\subsection{Group Maximum Fame Forecasting}

We have proposed two models to illustrate news generations:
log-normal model and HMM model. Now we will use a real forecasting
problem to examine the goodness of the two models.

Given the top 50 biggest cities (based on population) in the United
States, what is the probability for each city that has the maximum
fame in the group? Here the sentiment window size is 1 day. For
example, what's the probability that Miami gains the maximum media
exposure among all cities on today? Now we use two approaches to
solve this problem and compare the results.

\subsubsection{Using HMM Model}
There are two states, {\em Normal} and {\em Peak} states, in our HMM
model. To train a HMM model, we need to detect and fit pulses for
each city's time series. For each city $i$, we can compute
probabilities $\beta_i$ and $\gamma_i$ mentioned in Section
\ref{HMMnews} from training data and then the transition matrix
\ref{transition} could be computed. If a city $i$ has the maximum
fame within the group, two constraints should be satisfied: 1) city
$i$ is in the {\em Peak} state; 2) all other cities should be either
in {\em Normal} state, or in {\em Peak} state but their peak
references are smaller than city $i$'s reference. Therefore, city
$i$'s probability to reach the maximum fame in this group could be
calculated by $Pr_M(F_i)=$
\begin{small}
\begin{align*} Pr(\mathcal {P}_{F_i})
\times  \prod_{j=1,j\neq i}^n {[1-Pr(\mathcal {P}_{F_j}) +
Pr(\mathcal {P}_{F_j}) \times Pr(FP_i>FP_j)]}
\end{align*}
\end{small}
in which $Pr(\mathcal {P}_{F_i})$ is the probability that city $i$
is in the {\em Peak} state, and $Pr(FP_i>FP_j)$ is the probability
that city $i$'s peak reference is greater than city $j$'s peak
reference. All these probabilities could be calculated after the HMM
models are trained and built.

\begin{table}
\scriptsize \centerline{
\begin{tabular}
{l||r|r|r|r} \hline City & Days$_M$ & Pr(Real) & Pr(LN) & Pr(HMM)
\\ \hline \hline
New York  &  955 & 0.5323 & 0.5164 & 0.4598 \\
Washington & 749& 0.4175 &0.4337& 0.3801 \\
New Orleans& 43 & 0.0240 &2.34E-12  & 0.0342 \\
Chicago &11 &0.0061 &9.31E-06  & 0.0331 \\
Detroit &10& 0.0056 &4.04E-07 &0.0291 \\
Los Angeles &5 & 0.0028& 7.31E-08& 0.0138 \\
Houston &5 &0.0028 &3.33E-11 &0.0056 \\
Denver & 5 &0.0026 &2.57E-15 &0.0040 \\
Boston & 2 &0.0011 &1.77E-10 &0.0078 \\
Miami &  2& 0.0011 &2.14E-11 &0.0088 \\
Pittsburgh & 2 &0.0011& 2.07E-19& 0.0028 \\
Philadelphia& 1  & 0.0005 &6.64E-14 &0.0037 \\ \hline
\end{tabular}}
\caption{Result comparison for the top 15 referred cities
(2005-2009), for their actual occurrences and probabilities of
maximum peak days, probabilities of peak days calculated by
log-normal model (Pr(LN)) and HMM model (Pr(HMM)).
\label{citi-prob}}
\end{table}

\subsubsection{Using Truncated Log-normal Model}
From previous sections, we have already known that entity $i$'s
logged frequency $F_i$ follows a distribution $F_i \sim N(\mu_i,
{\sigma_i}^{2})$, in which the parameters $\mu_i$ and $\sigma_i$
could be computed from historical training data. If city $i$ gains
the biggest fame among all cities in this group, that means city $i$
should be more famous than any other ones. Then we have
\begin{align*}
Pr_M(F_i)&=Pr(F_i>F_1,F_i>F_2,...,F_i>F_n|i\neq j) \\
&=\prod_{j=1,j\neq i}^n {Pr(F_i>F_j)}
\end{align*}
$Pr_M(F_i)$ is the probability that city $i$ have the maximum fame
within this group, and $Pr(F_i>F_j)$ is the probability that city
$i$ is more famous than city $j$. $Pr(F_i>F_j)$ could be computed by
Monte Carlo simulation because we know both $F_i \sim N(\mu_i,
{\sigma_i}^{2})$ and $F_j \sim N(\mu_j, {\sigma_j}^{2})$.

\fullversion{
\begin{align}
  x^2 + y^2 &=  z^2 \\
  x^3 + y^3 &<  z^3
\end{align}

\begin{flalign}
  x^2 + y^2 &=  z^2 \\
  x^3 + y^3 &<  z^3
\end{flalign}
}

\subsubsection{Backtesting and Result Comparison}

We define city $i$'s ``peak day" as the days that city $i$ has the
maximum fame among all the cities in this group. Table
\ref{citi-prob} gives the result comparisons from 2005 to 2009
regarding the number of real peak days, real probabilities, and
probabilities calculated by log-normal and HMM models. We can see
the HMM method is much better than log-normal method because the
latter underestimated the probabilities of smaller cities
significantly.

An interesting phenomenon is that, cities' fame in news is not
equivalent to their sizes in terms of population. For example,
Washington, DC ranks 23th by population, but it is the second
popular city in news stream.

\section{Entity Fame Forecasting} \label{entityForecasting}

\begin{table*}[htp]
\scriptsize \centerline{
\begin{tabular}
{l|l||l|l|l|l|l||l|l|l|l} \hline \multirow{2}{*}{Pre Freq} &
\multirow{2}{*}{Set Size} & \multicolumn{5}{c||}{News Data} &
\multicolumn{4}{c}{Power Law Model} \\ \cline{3-11} & &
Cnts&MaxRef&MinRef&AveRef&Prob & slope & y-intercept & Cnts\_M &
Prob\_M
\\ \hline \hline
0-20&1022651&8&6736&3153&4388&7.82E-06&-1.415&0.079&14&1.44E-05\\
20-50&106540&4&12959&3703&6437&3.75E-05&-1.820&1.716&3&2.40E-05\\
50-100&53153&11&14242&3017&5651&2.07E-04&-2.313&4.218&8&1.50E-04\\
100-200&28280&22&7007&3007&4452&7.78E-04&-2.287&4.761&18&6.44E-04\\
\hline
\end{tabular}}
\caption{For entities with difference range of previous frequency
levels, we show the probabilities that they become famous, and the
maximum, minimum, and average references while becoming famous. Here
both the historical fame window  $w_m$ and the future fame window
$w_f$ are 1-day. We compare the probabilities computed from both
real news and our power-law models, and the results indicate our
models are pretty accurate. \label{toFamousPerformance}}
\end{table*}

\fullversion{

\begin{table*}
\footnotesize \centerline{
\begin{tabular}
{l|l|l|l|l} \hline Entity Name & Freq & Description & Why become
famous & On When
\\ \hline \hline
Mike Duvall&2932.8& politician& extramarital scandal exposed & 09/08/2009\\
Steve Irwin&2444& naturalist/zoologist& death & 09/04/2006\\
Eartha Kitt&2062& actress/singer& death & 12/25/2008\\
Steve McNair&1704& football player& death & 07/04/2009\\
Bobby Fischer&1625.2& chess player& death & 01/17/2008\\
Richard Widmark&1546.2& actor& death & 03/24/2008 \\
Tiffany Hall&1481.4& unknown woman & killed her friend's three young children & 09/15/2006\\
Michael Crichton&1353.4& author/producer& death & 11/04/2008\\
Kirsten Gillibrand&1338& politician& elected to the Senator of New York & 01/23/2008\\
Jim Cramer&1273& journalist/investor& interviewed by Jon Stewart on The Daily Show & 03/12/2009\\
Andrew Meyer&1180.2& university student& University of Florida Taser incident & 09/17/2007 \\
Victoria Osteen&1128.4& wife of Joel Osteen&  rejection of her previous Assault Claim & 08/13/2008\\
Trevor Immelman&1090 & golfer& win the 2008 Masters Tournament & 04/13/2008\\
Joe Andrew&1069 & author/investor& switch of endorsement from Hillary to Obama & 05/01/2008\\
Robert Rauschenberg&1037.6 & artist& death & 05/12/2008\\ \hline
\end{tabular}}
\caption{Some examples of entities which are from unknown to famous.
Before they became famous, their daily references are less than 20.
\label{fromUnknownToFamous}}
\end{table*}

}

\begin{table*}[htp]
\scriptsize \centerline{
\begin{tabular}
{l|l|l|l|l} \hline Entity Name & Freq & Description & Why become
famous & On When
\\ \hline \hline
Mike Duvall&2932.8& politician& extramarital scandal exposed & 09/08/2009\\
Steve Irwin&2444& naturalist/zoologist& death & 09/04/2006\\
Eartha Kitt&2062& actress/singer& death & 12/25/2008\\
Bobby Fischer&1625.2& chess player& death & 01/17/2008\\
Tiffany Hall&1481.4& unknown woman & killed her friend's three young children & 09/15/2006\\
Andrew Meyer&1180.2& university student& University of Florida Taser incident & 09/17/2007 \\
Trevor Immelman&1090 & golfer& win the 2008 Masters Tournament & 04/13/2008\\
Joe Andrew&1069 & author/investor& switch of endorsement from
Hillary to Obama & 05/01/2008\\ \hline
\end{tabular}}
\caption{Some examples of entities which are from unknown to famous.
Before they became famous, their daily references are less than 20.
\label{fromUnknownToFamous}}
\end{table*}

In the previous sections, we have studied news statistical patterns
and news generation models. In the following two sections, we will
build predictive models to predict future frequencies or fame of
news entities. This section focuses on entities' fame while the next
section focuses on the change rate of entities' fame, particularly
in a group-based context.

\begin{figure}[htp]
\centering
\includegraphics[width=1.04\linewidth]{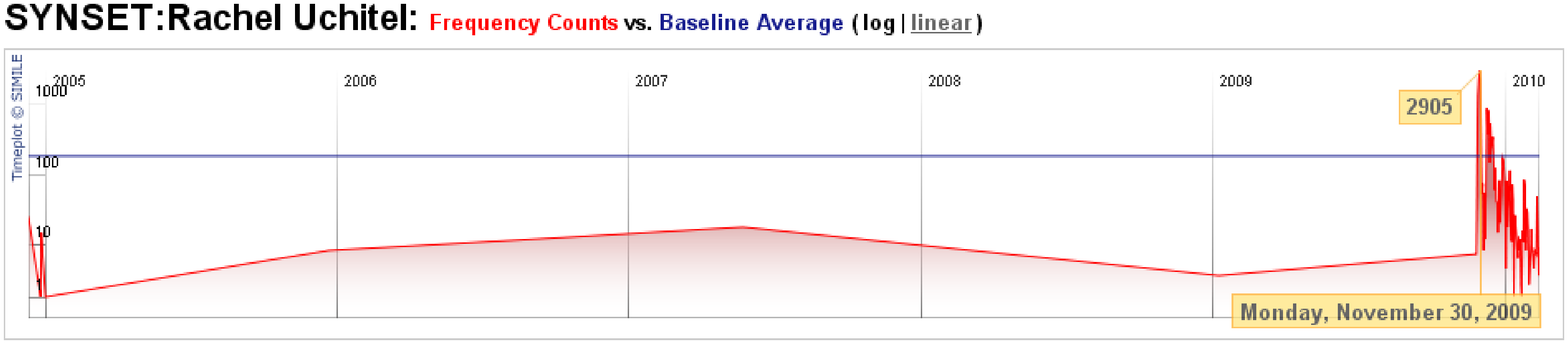}
\caption{Rachel Uchitel's time series. } \label{rachel}
\end{figure}

\begin{figure}[ht]
\centering
\includegraphics[width=0.49\linewidth,height=0.45\linewidth]{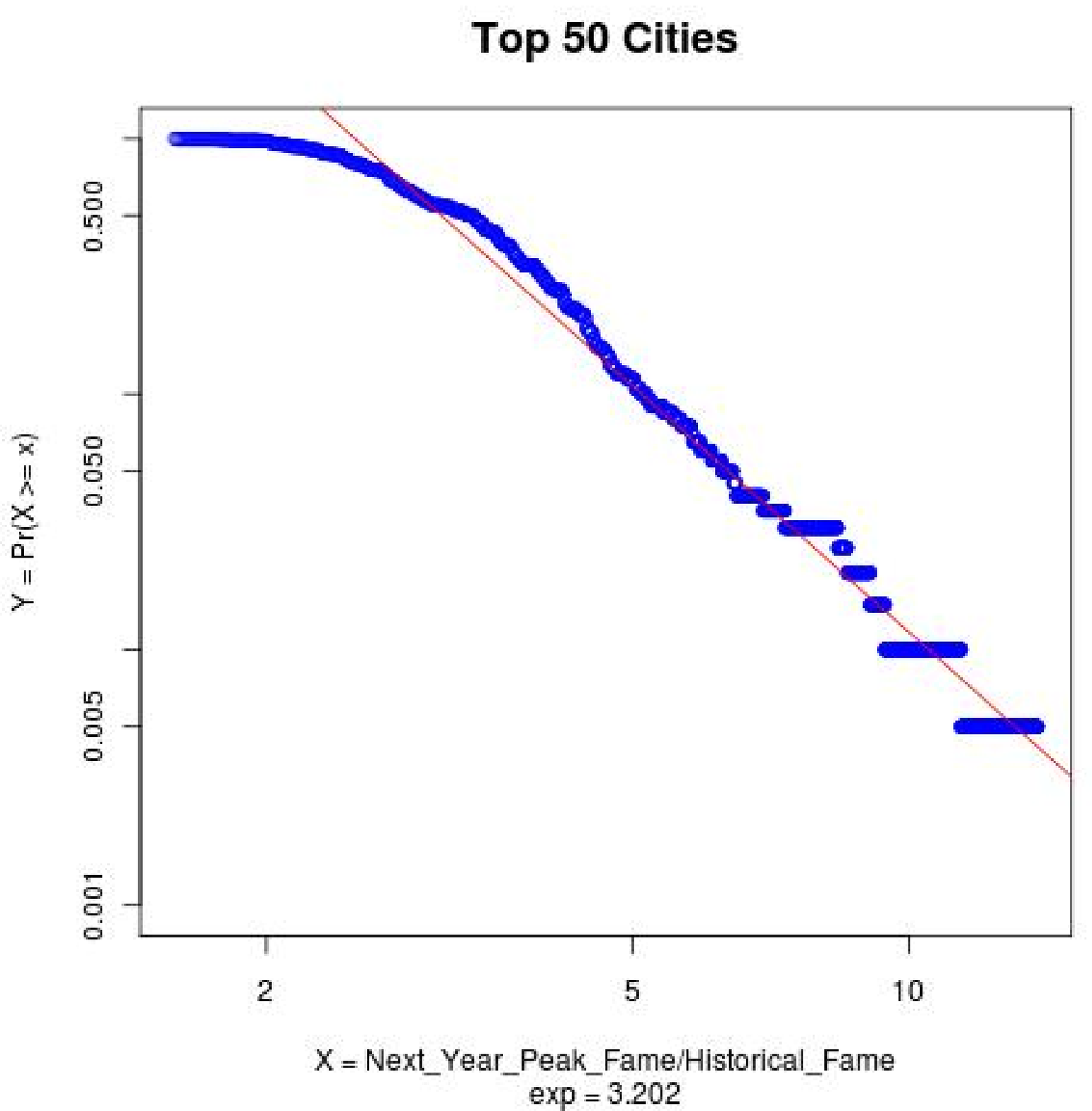}
\hfill
\includegraphics[width=0.49\linewidth,height=0.45\linewidth]{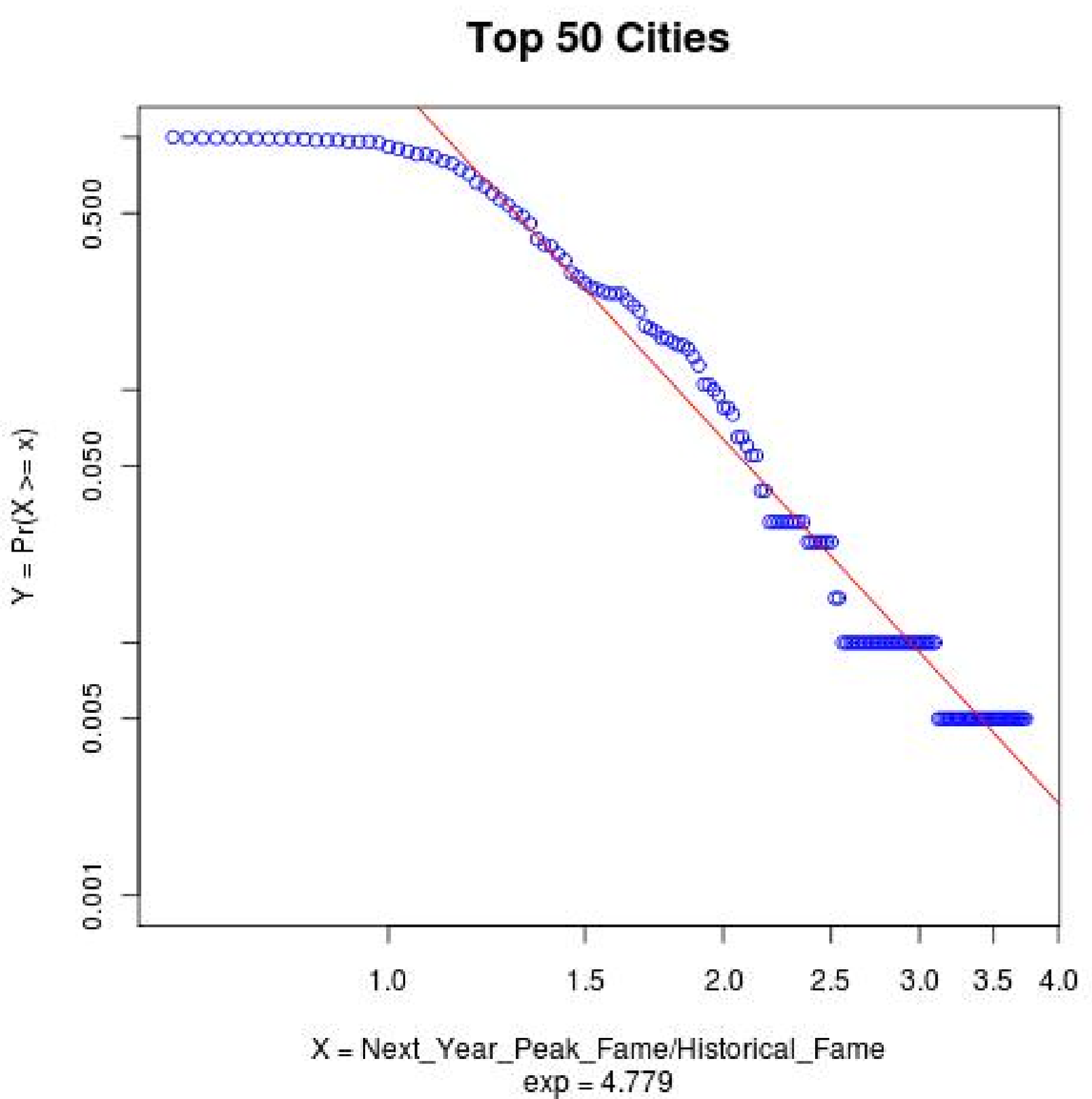}
\\
\includegraphics[width=0.49\linewidth,height=0.45\linewidth]{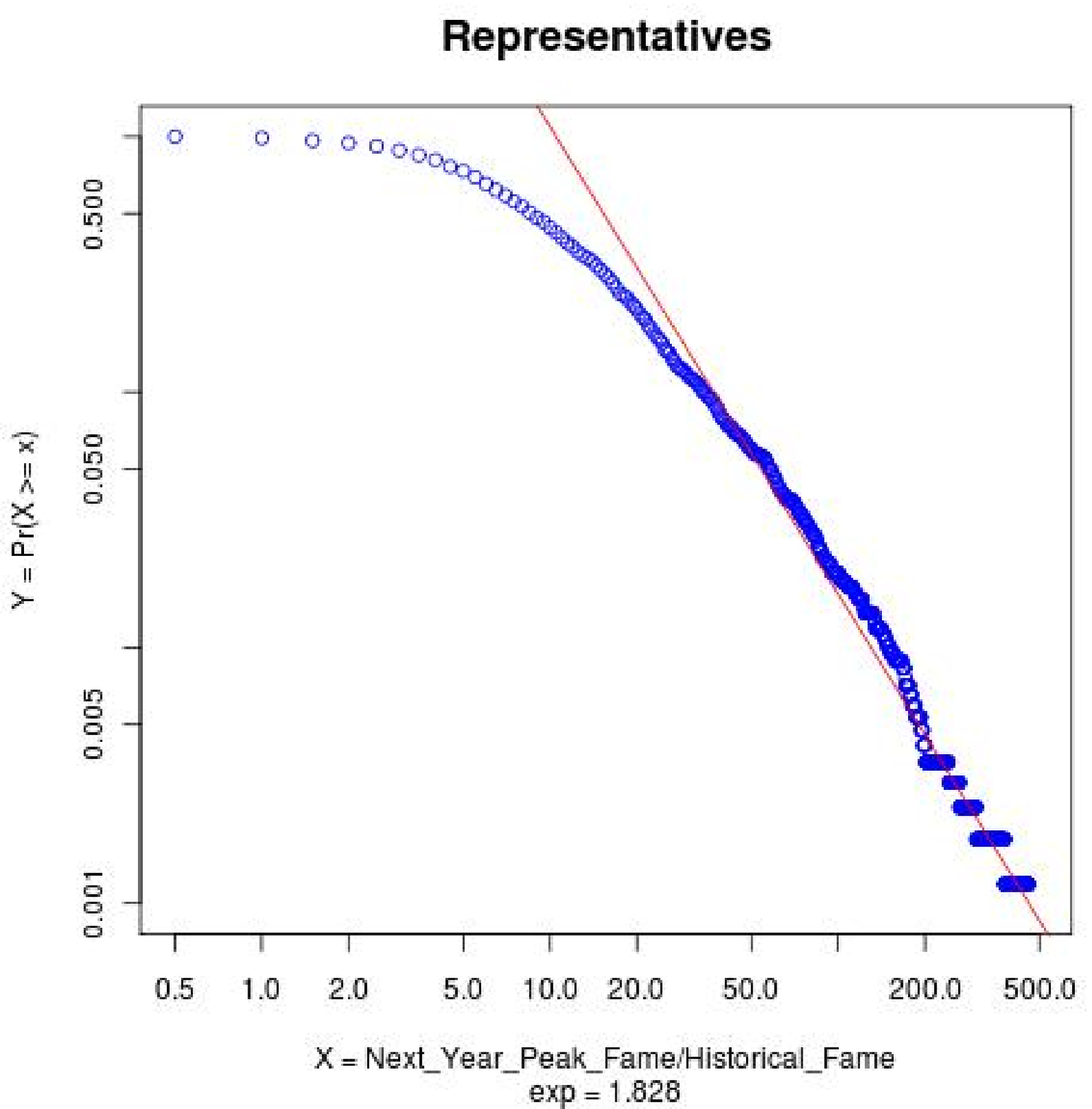}
\hfill
\includegraphics[width=0.49\linewidth,height=0.45\linewidth]{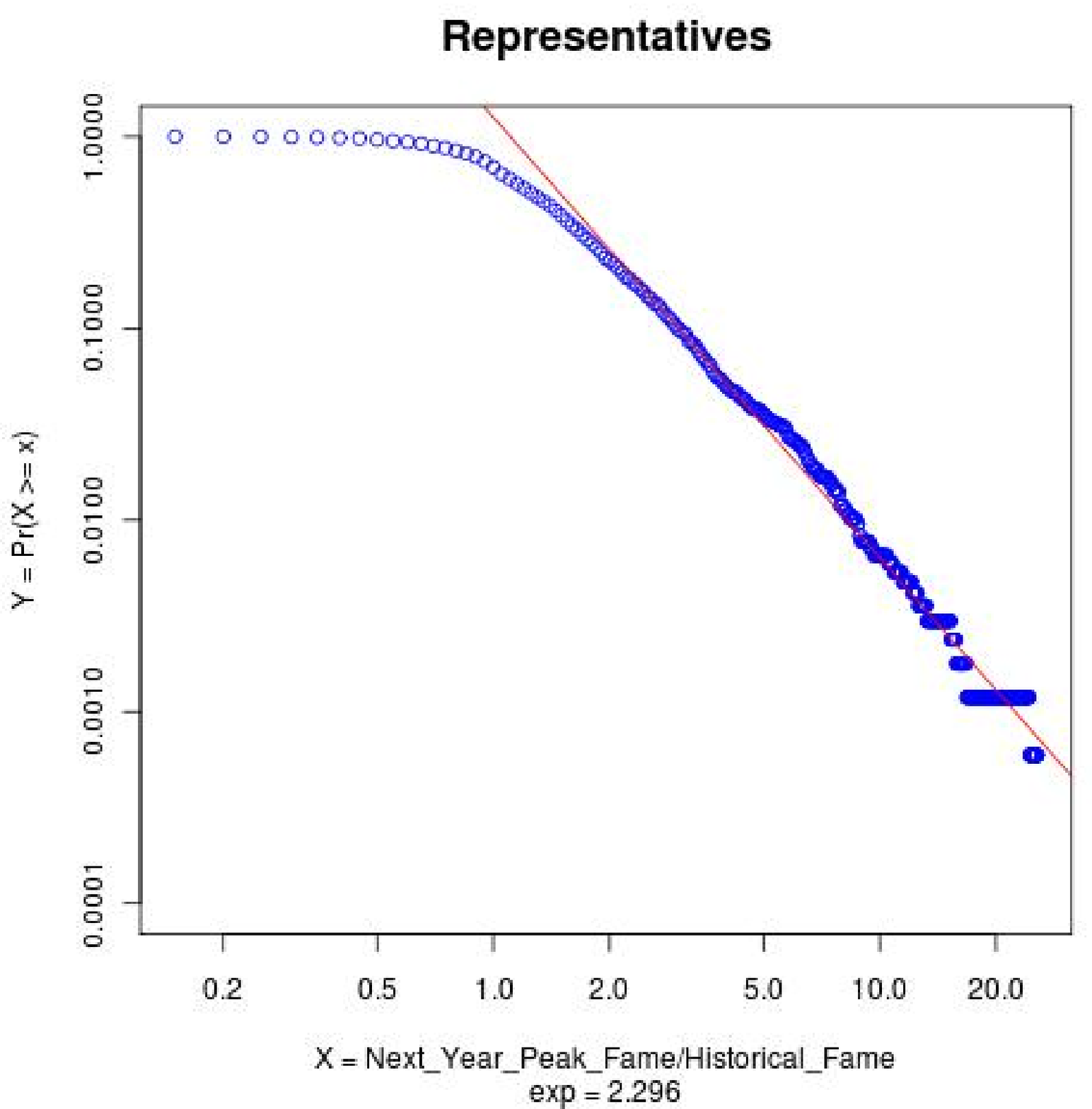}
\\
\includegraphics[width=0.49\linewidth,height=0.45\linewidth]{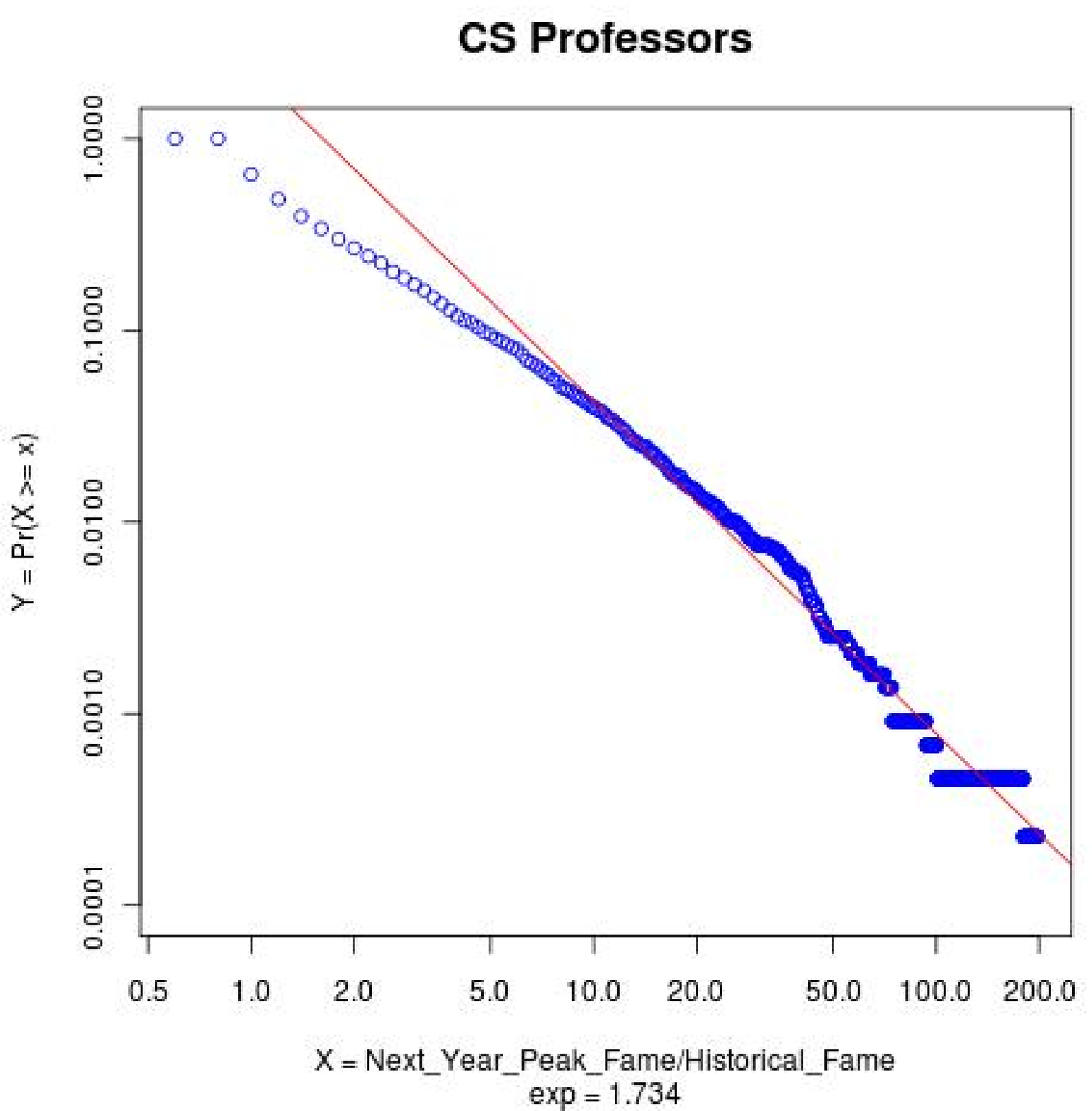}
\hfill
\includegraphics[width=0.49\linewidth,height=0.45\linewidth]{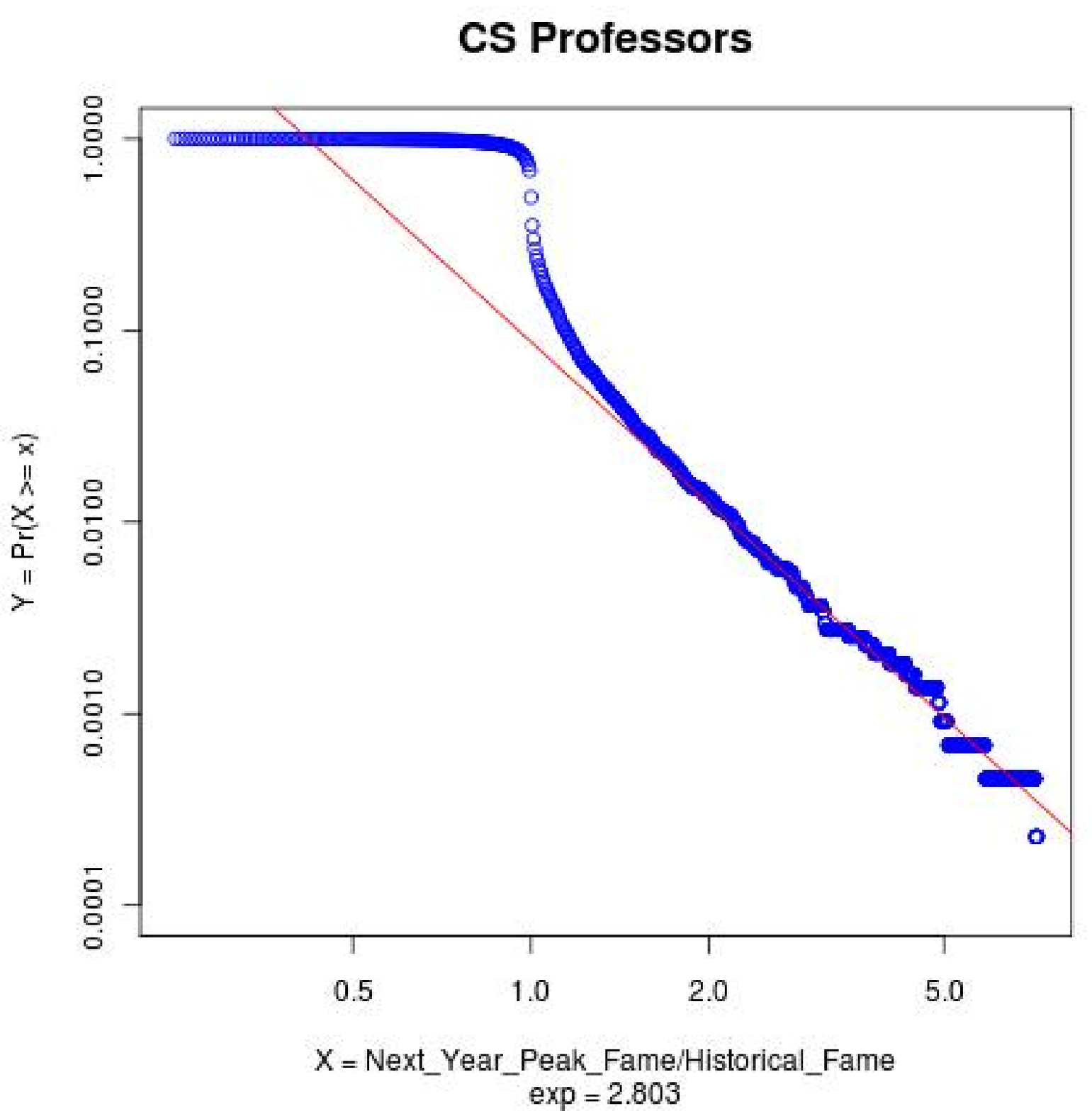}
\caption{Log-Log plot of $R_{ph}(G, t)$ (left column) and $R_{ah}(G,
t)$ (right column) for groups {\em Top 50 cities}, {\em
Representatives}, and {\em CS professors} respectively. Here $t$ is
1 year. The x-axis is the ratio $x$ ($R_{ph}$ or $R_{ah}$), and the
y-axis is $Pr(X >x)$. } \label{groupForecast}
\end{figure}

Probably the most interesting forecasting questions is, how can we
forecast the probabilities that some unknown news entities become
extremely famous? For example, Rachel Uchitel (Figure \ref{rachel})
appeared in news since the very beginning, but she only had very
little fame and kept quite until the end of November 2009 because of
Tiger Woods' sex scandal. She became well known since that time. Our
key question is how to estimate the probability that she becomes
famous.

For a certain group $G$, if we know today's fame for each entity,
what is the forward distribution of tomorrow's fame? For example,
for all entities currently with daily references in range (0,20),
how their tomorrow's frequencies $f(G, 0, 20)$ are distributed? More
generally, we denote $m_l$ and $m_u$ are the lower and upper bounds
of historical frequencies/fame. If current day's fame (with fame
window size $w_m$) is in range $(m_l, m_u)$, how tomorrow's fame
(with fame window size $w_f$)is distributed? Here we can denote the
fame as $f(G, m_l, m_u, w_m, w_f)$. In addition, we define that
notion $N(G, m_l, m_u, w_m, w_f)$ is the probability distribution of
that frequency/fame is greater than some certain level $x$. That is,
the distribution is $F_i \sim N(G, m_l, m_u, w_m, w_f) = Pr(X>x)$.
Figure \ref{forwardDist} shows that both the histogram plot of $f(G,
m_l,$ $m_u, w_m, w_f)$ and the distribution $N(G, m_l, m_u, w_m,
w_f)$ have power-law tails, in spite of their different
historical/future fame window sizes. However, the histogram plots
\ref{fig:b},\ref{fig:c}, and \ref{fig:d} are somewhat truncated into
two parts by certain truncation points. The truncation points are
some values between $ln(m_l)$ and $ln(m_u)$. If $m_l = m_u$, the
truncation value is exactly $ln(m_l)$, just as the truncation point
shown in Figure \ref{fig:c}, with a value of $ln(30) = 3.4$.

Now Using a subset of our news database as the training data, we
will apply this model to estimate the probabilities that some
trivial entities become very famous. The results are shown in Table
\ref{toFamousPerformance}. Here both the historical fame window size
and the future fame window size is set to 1 day. This table shows
that some unknown entities with little fame (historical references
0$\sim$200) became famous (future references $> 3000$). Actually,
Table \ref{entityFame} tells us that a reference of 3000 means a
similar fame with Chicago. We divide entities into four categories
according to their historical fame, (0-20), (20-50), (50-100), and
(100-200) respectively, and then we train the data to get the slopes
and y-intercepts to build four power-law models. We compare the
entity counts and probabilities to become famous between real news
data and our power-law models, and find they are very close. For
example, while $m_l=0$ and $m_u=20$, we have
\begin{equation}
log P = -1.415*log (x) +0.079
\end{equation}
To make $x=3000$, we  get $P = 1.44E-05$ and the estimated counts to
become famous is $P*1022651\approx 14$.

\fullversion{ Our model also shows, for unknown entities with
average daily references below 20, we have  probability 7.82E-06
that make them famous, which means these entities will have a daily
reference above 3000; and have probability 9.77E-07 that make them
extremely famous, which means these entities will have a daily
reference above 6000. Therefore, in order to make an unknown entity
famous, we need to hold $128K$ news entities and wait for 1 day, or
hold 350 entities for 1 year, or hold 1 entity for 359 years.
Similarly, in order to make an unknown entity extremely famous, we
need to hold 1 million entities for 1 day, or hold 2804 entities for
1 year, or hold 1 entity for 2804 years.
}

The final question is that what kind of entities become famous.
Table \ref{fromUnknownToFamous} lists some unknown entities (with
daily reference $<$ 20) became famous (with daily reference $>$ 1000
for 5 continuous days). We can see many of them became famous
because of death, and some of them were because of political events.
This tells us the interesting fact that media tends to remember
people accompanied with their death. Some non-people entities could
also become famous, e.g., ``Sichan" and ``Haiti" became famous
because of earthquakes.

\section{Group-based Fame Forecasting}
\label{groupForecasting}

Now we consider the fame change of entities within a specific domain
or group context. The fact that an entity becomes famous means its
fame changes dramatically, which indicates its future fame is
significantly higher than its historical fame. For entities in a
group $G$ over a future time range $t$, the degree of fame-change
could be measured by ratio $R_{ph}(G, t)$ or $R_{ah}(G, t)$ shown as
below:
\begin{list}{\labelitemi}{\leftmargin=1.5em} \itemsep -1pt
\item $\displaystyle R_{ph}(G, t) = \frac{next\ period\ (t)\ peak\ fame}{historical\ fame} = \frac{p(t, w_f)}{h}$
\item $\displaystyle R_{ah}(G, t) = \frac{next\ period\ (t)\ average\ fame}{historical\ fame} = \frac{a(t)}{h}$
\end{list}
Next period peak fame is a function of future time range $t$ and
peak fame window size $w_f$. In our analysis, we make $t$ as 1 year
and $w_f$ as 5 days.

Figure \ref{groupForecast} indicates that the distributions of both
$R_{ph}(G, t)$ and $R_{ah}(G, t)$ have power-law tails, although the
slopes in their Log-Log plots are not exactly the same for different
groups. The fitted linear lines are also shown in these Log-Log
plots. Based on the power-law models, we can compute the probability
that some entity within the group becomes famous in a future time
range $t$. We have
\begin{equation}
Pr(X>x)=cx^{-\lambda},\ for\ x\geq x_{min}
\label{groupPowerLawFormula}
\end{equation}
That is $logP= -\lambda *log(x) + log(c)$. By fitting $\lambda$ and
$log(c)$ with a linear model, $P$ could be calculated. Table
\ref{groupModelComparison} examines the accuracy of the power-law
model, which indicates the theoretical result and the real news data
match very well.

Table \ref{groupFromUnknownToFamous} gives some examples that
entities have big values of $R_{ph}(G, t)$ or $R_{ah}(G, t)$. Indeed
we can use distribution either $R_{ph}(G, t)$ or $R_{ah}(G, t)$ to
calculate probabilities that entities become famous, but we may get
slight different results with these two ratios.

\begin{table*}[htp]
\scriptsize \centerline{
\begin{tabular}{
l|l||l|l|l|l|l|l||l|l|l|l|l|l} \hline \multirow{3}{*}{Groups} &
\multirow{3}{*}{Size} &
\multicolumn{6}{c||}{Next\_Year\_Peak\_Fame/Historical\_Fame} &
\multicolumn{6}{c}{Next\_Year\_Ave\_Fame/Historical\_Fame} \\
\cline{3-14} && & \multicolumn{2}{c|}{Real Data}&
\multicolumn{3}{c||}{Our Model} & &
\multicolumn{2}{c|}{Real Data}& \multicolumn{3}{c}{Our Model} \\
\cline{3-14} && T &Cnts&Prob&Slope&Cnts&Prob&T&Cnts&Prob&Slope&Cnts&Prob \\
\hline \hline
\multirow{3}{*}{Top Cities} & \multirow{3}{*}{50} &5&16&0.080&-3.202&21&0.107&2&17&0.085&-4.779&13&0.064 \\ &&7&7&0.035&-3.202&7&0.036&2.5&5&0.025&-4.779&4&0.022 \\
&&10&2&0.010&-3.202&2&0.010&3&2&0.01&-4.779&2&0.009 \\
\hline \hline
\multirow{3}{*}{Representatives} & \multirow{3}{*}{439} &50&101&5.75E-02&-1.828&101&5.76E-02&5&60&3.41E-02&-2.296&55&3.13E-02 \\ &&100&33&1.88E-02&-1.828&28&1.62E-02&10&11&6.26E-03&-2.296&11&6.37E-03 \\
&&200&7&3.99E-03&-1.828&8&4.57E-03&20&2&1.14E-03&-2.296&2&1.29E-03 \\
\hline \hline
\multirow{3}{*}{CS Profs} & \multirow{3}{*}{1911} & 50 & 11 &1.44E-03&-1.734&19&2.60E-03&3&16&2.09E-03&-2.803&30&4.00E-03 \\ &&100&3&3.92E-04&-1.734&5&7.82E-04&5&4&5.23E-04&-2.803&7&9.56E-04 \\
&&200&0&0&-1.734&2&2.34E-04&7&2&2.61E-04&-2.803&3&3.72E-04 \\
\hline
\end{tabular}}
\caption{Accuracy of our models, with the result from a small group
(top 50 cities), a medium group (representatives), and a large group
(CS professors) respectively. We evaluated both $R_{ph}$ and
$R_{ah}$ ratios. The two ``T"-columns are thresholds, and the counts
and probabilities evaluate the possibilities that $R_{ph}$ or
$R_{ah}$ is greater than the thresholds. Estimations from our
power-law model are very close to the real news.
\label{groupModelComparison}}
\end{table*}

\begin{table*}
\scriptsize \centerline{
\begin{tabular}
{l|l|r|r|r|l|l} \hline Entity Name & Group & P/H & A/H & Peak Freqs
& Why become famous & On When
\\ \hline \hline
New Orleans & Cities & 10.23 & 1.30& 24917 & Hurricane Katrina &
09/02/2005 \\ Memphis & Cities & 12.39& 2.54 & 7763 & the first team
in NCAA to achieve 30 wins in a season & 04/07/2008
\\
Omaha &Cities & 8.55 & 3.11 & 2622 &  June 2008 tornado outbreak
sequence & 06/12/2008 \\
Tulsa & Cities & 9.64&3.76 & 5563 & ice storm in December &
12/26/2007 \\ \hline Kirsten Gillibrand & Represen & 458.8
&24.2&5324 & elected to the Senator of New York & 01/23/2009 \\
Joe Wilson & Represen & 491.7 & 13.2 & 23323 & shouted at Obama during 2009 Presidential address & 09/09/2009 \\
\hline Sebastian Thrun & CS Profs &
74.3& 3.1& 1091 & helps GM to make robot driven cars & 01/07/2008\\
Amar Bose
& CS Profs & 51.2& 3.7 & 652 &retired from MIT & 11/26/2005 \\
\hline
\end{tabular}}
\caption{Some examples of entities with high $R_{ph}$ or $R_{ah}$
ratios. \label{groupFromUnknownToFamous}}
\end{table*}

\begin{list}{\labelitemi}{\leftmargin=1em} \itemsep -1pt
\item {\bf Probability of making zero-fame people famous}
\end{list}

Based on the power-law tail, we can estimate what's the probability
that some people with almost zero fame became famous, e.g., as
famous as Tiger Woods. From Table \ref{entityFame}, we know Tiger
Woods' peak fame is 20,000, and we notice the power-law model
($R_{ph}$) for CS professors (let's assume this model is generally
applicable to any group of people) is
\begin{equation}
log P = -1.734*log X +0.362
\end{equation}
While $X=20,000$, we can get $P=8.017E-08$. We know there are 300
million people in the United States. Therefore, there are roughly
$P*300$ million = 24 persons that will have comparable peak fame
with Tiger Woods in the next 1 year.

But how about to reach Tiger Woods' average fame? We know the
power-law model ($R_{ah}$) for CS professors is
\begin{equation}
log P = -2.803*log X -1.060
\end{equation}
Because Woods' average fame is around 1000, our calculation shows
there is only 0.1 person who has no previous fame at all but can
reach Tiger Woods' average fame in the next year. Similarly, Steve
Jobs' average fame is around 100, and our model shows there are
roughly 64 unknown persons can reach Steve's average fame in the
next year.

In all previous cases, we make $t$ as 1 year to train power-law
models. But how about 10 years? That is, what is probability of an
unknown person become famous in the next 10 years? Let's suppose the
distribution of $R_{ph}(G, t)$ or $R_{ah}(G, t)$ follows Formula
\ref{groupPowerLawFormula} for future time range $t$. Now we could
deduce the distribution for a time range of $n\times t$:
\begin{align*}
Pr_n(X>x)&=1-Pr_n(X\leq x)=1- \prod_{i=1}^n {Pr_1(X_i\leq x)}\\
&=1-(1-cx^{-\lambda})^n \approx 1-(1-ncx^{-\lambda}) \\
&=n\times cx^{-\lambda}
\end{align*}
So we argue that the probability just increases linearly with the
increasing of time.

\section{Conclusions} \label{conclusions}

This paper studied new entity and group modeling and forecasting
methodologies, including group fame distribution analysis, group
fame probability analysis, and group fame evolution over time. We
show some important news entity and group statistical patterns could
be described by log-normal or power-law distributions. We also
proposed a HMM-based news generation model, which has never been
used in news modeling before. We show that HMM models are more
capable of describing news generations than simple Log-Normal
models. Based on these analysis, we answered some interesting news
forecasting questions. For example, what is the probability that an
entity become the most famous one among its group? And what is the
likelihood that a trivial entity becomes incredibly important in the
next time period? Our analysis shows these questions could be solved
by fitting power-law tails and we validated the model with several
interesting news groups in different domains. Our study provides
very useful insights for the analysis of issues in finance,
political science, or social science.

%
\bibliographystyle{abbrv}
\bibliography{refs}  
%
%


\end{document}